\documentclass[
twocolumn, %preprint,
superscriptaddress,
 showpacs,preprintnumbers,
 amsmath,amssymb,
 aps,
 prb,
]{revtex4-1}
\usepackage{graphicx}
\usepackage{dcolumn}% Align table columns on decimal point
\usepackage{bm}% bold math
\begin{document}

\title{Noninvasive Embedding of Single Co Atoms in Ge(111)2$\times$1 Surfaces}

\author{D.~A.~Muzychenko}
\email{mda@spmlab.ru} \affiliation{Faculty of Physics, Moscow State University, 119991 Moscow, Russia}
\author{K.~Schouteden}
\email{Koen.Schouteden@fys.kuleuven.be} \affiliation{Laboratory of Solid-State Physics and Magnetism, KULeuven, BE-3001 Leuven, Belgium}
\author{M.~Houssa}
\affiliation{Semiconductor Physics Laboratory, Department of Physics and Astronomy, KULeuven, BE-3001 Leuven, Belgium}
\author{S.~V.~Savinov}
\affiliation{Faculty of Physics, Moscow State University, 119991 Moscow, Russia}
\author{C.~Van~Haesendonck}
\affiliation{Laboratory of Solid-State Physics and Magnetism, KULeuven, BE-3001 Leuven, Belgium}

\date{\today}

\begin{abstract}
We report on a combined scanning tunneling  microscopy (STM) and density functional theory (DFT) based investigation of Co atoms on
Ge(111)2$\times$1 surfaces. When deposited on cold surfaces, individual Co atoms have a limited diffusivity on the atomically flat areas and
apparently reside on top of the upper $\pi$-bonded chain rows exclusively. Voltage-dependent STM imaging reveals a highly anisotropic electronic
perturbation of the Ge surface surrounding these Co atoms and pronounced one-dimensional confinement along the $\pi$-bonded chains. DFT
calculations reveal that the individual Co atoms are in fact embedded in the Ge surface, where they occupy a quasi-stationary position within
the big 7-member Ge ring in between the $3^{\rm rd}$ and $4^{\rm th}$ atomic Ge layer. The energy needed for the Co atoms to overcome the
potential barrier for penetration in the Ge surface is provided by the kinetic energy resulting from the deposition process. DFT calculations
further demonstrate that the embedded Co atoms form four covalent Co--Ge bonds, resulting in a Co$^{\rm 4+}$ valence state and a 3$d^{\rm 5}$
electronic configuration. Calculated STM images are in perfect agreement with the experimental atomic resolution STM images for the broad range
of applied tunneling voltages.
\end{abstract}



\maketitle

%%%%%%%%%%%%%%%%%%%%%%%%%%%%%%%%%%%%%%%%%%%%%%%%%%%%%%%%%%%%%%%%%%%%%%%%%%%%
%%%%%%%%%%%%%%%%%%%%%%%%%%%%%%%%%%%%%%%%%%%%%%%%%%%%%%%%%%%%%%%%%%%%%%%%%%%%
%%%%%%%%%%%%%%%%%%            Introduction                  %%%%%%%%%%%%%%%%
%%%%%%%%%%%%%%%%%%%%%%%%%%%%%%%%%%%%%%%%%%%%%%%%%%%%%%%%%%%%%%%%%%%%%%%%%%%%
%%%%%%%%%%%%%%%%%%%%%%%%%%%%%%%%%%%%%%%%%%%%%%%%%%%%%%%%%%%%%%%%%%%%%%%%%%%%
\section{Introduction}
The continuous miniaturization of electronic circuits has resulted in the emergence of novel classes of nanometer size devices that rely on the
quantum-mechanical nature of charge carriers.~\cite{Bhattacharya_AnnRev_04, Kuo_Nature_05} Examples of state-of-the-art nanodevices can be found
in molecular electronics~\cite{Whalley_JAmChemSoc_07, Grigoriev_PRB_06, Parks_Science_10} and spintronics.~\cite{Prinz_PT_95} The dimensions of
the active elements, connections and separations are now being reduced to the order of a few atomic rows and, in the ultimate limit, devices may
be built up using atomic size elements~\cite{Haider_PRL_09, Tan_NanoLett_09} that are connected by atomic nanowires. However, to fulfill the
demands related to the never ceasing development of electronics, novel materials with electronic properties superior to that of the currently
used silicon are required. Among all candidates germanium is considered as one of the most promising alternative materials~\cite{Bracht_PRL_09,
Wundisch_APL_09} because it allows higher switching speeds due to a lower effective hole mass and a higher electron and hole drift
mobility.~\cite{Sze_SolStEl_68} This makes germanium ideally suited for use in ultrafast complementary metal-oxide-semiconductor technology, in
particular for metal-oxide-semiconductor field-effect transistors~\cite{Minjoo_JAP_05, Leitz_JAP_02, Claeys_Simoen_07} and band-to-band
tunneling field-effect transistors.~\cite{Appenzeller_PRL_04} For this purpose detailed investigations of the electronic properties of
dopants/defects or metal alloys in Ge crystals are obviously required.~\cite{Chui_APL_03}

One of the major challenges for future nanoelectronic applications is the controlled preparation of low-dimensional structures on semiconductor
surfaces, e.g. quantum dots~\cite{Reimann_RMP_02} and quantum wires.~\cite{Nilius_Sci_02, Wang_PRB_04, Schouteden_NT_09} Due to their broad
range of electronic and magnetic properties, such nanostructures are ideal model systems for the fundamental study of low-dimensional physics as
well as for the exploration of new device concepts~\cite{Kouwenhoven_RPP_01} that also exploit the spin character of the charge
carriers.~\cite{Prinz_PT_95, Wolf_Sci_01, Zutic_RMP_04}

Within this context, deposition of metal atoms on Ge surfaces has attracted considerable scientific interest during recent years, since it was
found that atoms of different materials self-organize into different types of nanostructures after deposition on Ge. It has been demonstrated
that deposited Mn atoms do not coagulate on Ge(111)\emph{c}2$\times$8 surfaces in the initial adsorption stage, yielding zero-dimensional (0D)
structures on Ge(111).~\cite{Changgan_PRB_04, Profeta_PRB_04} On the other hand, Pt,~\cite{Oncel_PRL_05, Schafer_PRB_06, Stekolnikov_PRL_08,
Stekolnikov_PRB_08, Vanpoucke_PRB_10} Au~\cite{Sauer_PRB_10, Houselt_PRB_08} and Sn atoms~\cite{Tomatsu_SS_07} spontaneously form
one-dimensional (1D) atomic chains on Ge(001), whereas Pd~\cite{Fischer_PRB_07, Fischer_PRB_07, Wang_JAP_06} and Ag~\cite{Lince_JVST_83,
Chan_PRB_02} atoms favor the formation of two-dimensional (2D) or three-dimensional (3D) particles on Ge(001).

The emerging field of spintronics requires (self)assembly of nanostructures with well-defined magnetic properties on semiconducting surfaces.
Due to the high spin polarization of the charge carriers near the Fermi level, Co is one of the most important elements used in magnetic
recording media as well as in giant magnetoresistance devices.~\cite{Parkin_PRL_93} Recently, the electronic and magnetic behavior of ultrathin
($\leq 5$ monolayers) Co/Ge~\cite{Tsay_SS_02, Ryan_PRB_04, Tsay_SS_04, Tsay_JVSTA_05, Chang_JAP_06} and Co/Ag/Ge~\cite{Tsay_APL_03,
Tsay_JMMM_04, Su_JMMM_06} films has been investigated. The initial adsorption stage of single Co atoms on Ge surfaces has not been studied so
far. Thorough knowledge of the formation process of the Co/Ge interface during the first adsorption stages is, however, of crucial technological
and fundamental interest.

Here, we present a comprehensive study of the initial growth stages of Co on 2$\times$1 reconstructed Ge(111) by means of low-temperature (LT)
scanning tunneling microscopy (STM) and spectroscopy (STS), combined with first-principles density functional theory (DFT) calculations within
the local-density approximation. STM and STS are ideal tools to investigate with high spatial and energy resolution the surface reconstruction
and the local electronic properties of the Ge(111) surface after adsorption of individual Co atoms. DFT calculations on the other hand allow us
to predict the electronic properties of systems of up to thousands of atoms in size. High resolution STM/STS combined with DFT hence provides a
powerful tool for the investigation of atomic size systems. Here, we report on the first experimental observation of ``noninvasive embedding''
(i.e. without destroying the surface reconstruction) of individual Co atoms in the Ge(111)2$\times$1 surface and on the formation of larger
Co/Ge intermixed layers after Co deposition on the cold ($T_{\rm sample} \leq  80 \, {\rm K}$) Ge(111)2$\times$1 surface. The location of an
individual Co atom in the Ge surface, its influence on the surrounding Ge atoms and the resulting electronic properties are systematically
investigated. Voltage-dependent STM imaging reveals a highly anisotropic electronic perturbation of the Ge surface surrounding the Co atom,
which is accompanied by pronounced 1D confinement along the $\pi$-bonded chains. Our experimental findings are well explained by the detailed
DFT calculations.

%%%%%%%%%%%%%%%%%%%%%%%%%%%%%%%%%%%%%%%%%%%%%%%%%%%%%%%%%%%%%%%%%%%%%%%%%%%%
%%%%%%%%%%%%%%%%%%%%%%%%%%%%%%%%%%%%%%%%%%%%%%%%%%%%%%%%%%%%%%%%%%%%%%%%%%%%
%%%%%%%%%%%%%%%%%%          Instrumentation                 %%%%%%%%%%%%%%%%
%%%%%%%%%%%%%%%%%%%%%%%%%%%%%%%%%%%%%%%%%%%%%%%%%%%%%%%%%%%%%%%%%%%%%%%%%%%%
%%%%%%%%%%%%%%%%%%%%%%%%%%%%%%%%%%%%%%%%%%%%%%%%%%%%%%%%%%%%%%%%%%%%%%%%%%%%
\section{Instrumentation}
\label{sect:Inst}

STM and STS measurements were performed with a LT STM setup (Omicron Nanotechnology), operating at a base pressure in the $10^{-11} \, {\rm
mbar}$ range. All data are acquired at $T_{\rm sample} \simeq 4.5 \, {\rm K}$. Electrochemically etched W tips were cleaned \emph{in situ} by
repeated flashing well above $1800 \, {\rm K}$ to remove the surface oxide layer and any additional contamination. The tip quality was routinely
checked by acquiring atomic resolution images of the ``herringbone'' reconstruction of the Au(111) surface.~\cite{Barth_JV_90,
Schouteden_NJP_08} STM topographic imaging was performed in constant current mode. The tunneling voltages $V_{\rm t}$ indicated in the text and
figure captions are with respect to the sample (the STM tip is virtually grounded). Image processing was performed by Nanotec WSxM.~\cite{WXsM}

Ge single crystals with a resistivity of $\rho_{\rm bulk} \simeq 0.2 \, {\rm \Omega cm}$ are doped with Ga at a doping level of $n_{\rm Ga} = 1$
to $2 \times 10^{16} \, {\rm cm}^{-3}$, resulting in $p$-type bulk conductivity. $4 \times 1.5 \times 0.8 \ \, {\rm mm}^3$ Ge bars, with their
long axis aligned with the (111) direction, were cleaved \emph{in situ} at room temperature in the sample preparation chamber at a pressure of
around $5 \times 10^{-11}\, {\rm mbar}$. The freshly cleaved samples were transferred within about 5 minutes to the STM measurement chamber. The
pressure in the STM measurement chamber was about $4 \times 10^{-12} \, {\rm mbar}$ during the LT STM measurements. Under these conditions the
cleaved Ge surfaces were observed to retain their cleanliness for 5 to 7 days. This way, we have investigated 7 freshly cleaved Ge(111)
crystals.

After checking the freshly cleaved Ge(111)2$\times$1 sample in the STM measurement chamber, $0.02 \, \rm{to} \, 0.04$ monolayers of Co are
deposited on the cold Ge(111)2$\times$1 surface ($T_{\rm sample} \leq 80 \, {\rm K}$) in the sample preparation chamber. Deposition was achieved
by evaporation from a high purity Co (99.9996\%) rod with an e-beam evaporator, at pressures below $10^{-10} \, {\rm mbar}$ and at a low
deposition rate of around $0.007 \pm 0.001$ monolayers (MLs) per seconds. After Co deposition, the Ge(111)2$\times$1 sample is transferred
immediately to the LT STM measurement chamber. Overall transport time, including deposition time, was around $25 \, {\rm minutes}$. Here we
focus on the results obtained on 4 different Co/Ge(111)2$\times$1 samples for which $2.1 \, {\rm \mu m}^{2}$ atomic resolution STM topography
images were recorded and analyzed.

%========================================================%
%==========   Fig 1     " 2x1_schematic "    ============%
%========================================================%
\begin{figure}
\includegraphics[width=84mm, scale=1.00]{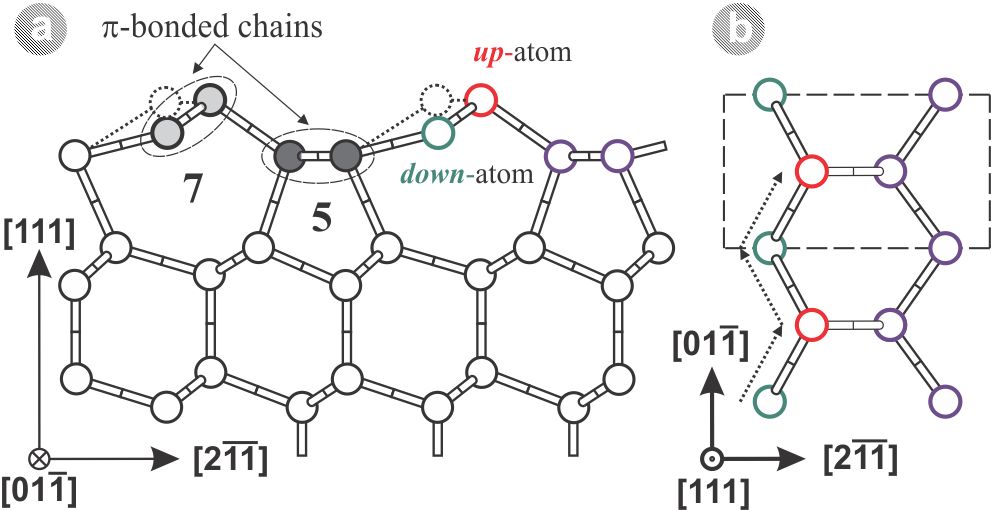}
\caption{(Color online) (a) Schematic side view of the chain-left isomer of the 2$\times$1 reconstruction of the Ge(111) surface according to
the original Pandey $\pi$-bonded chain model (dotted lines) and including the effect of buckling (solid lines).~\cite{Northrup_PRB_83} The
7-member and 5-member Ge rings of the surface reconstruction are indicated by the numbers 7 and 5, respectively. (b) Schematic top view of
``zigzag'' chain structure of the three top layers in (a). The dashed frame indicates the surface unit cell, while the arrows with dotted lines
indicate the ``zigzag'' structure of the upper $\pi$-bonded chain along $[01\overline{1}]$ direction.} \label{2x1_schematic}
\end{figure}
%========================================================%
%========================================================%
%========================================================%

%%%%%%%%%%%%%%%%%%%%%%%%%%%%%%%%%%%%%%%%%%%%%%%%%%%%%%%%%%%%%%%%%%%%%%%%%%%%
%%%%%%%%%%%%%%%%%%%%%%%%%%%%%%%%%%%%%%%%%%%%%%%%%%%%%%%%%%%%%%%%%%%%%%%%%%%%
%%%%%%%%%%%%%%%%%%         Experimental Results             %%%%%%%%%%%%%%%%
%%%%%%%%%%%%%%%%%%%%%%%%%%%%%%%%%%%%%%%%%%%%%%%%%%%%%%%%%%%%%%%%%%%%%%%%%%%%
%%%%%%%%%%%%%%%%%%%%%%%%%%%%%%%%%%%%%%%%%%%%%%%%%%%%%%%%%%%%%%%%%%%%%%%%%%%%
\section{Experimental Results}
\label{sect:Exp_Res}

%--------------------------------------------------------------------------%
\subsection{Topography and electronic structure of the freshly cleaved Ge(111)2$\times$1 surface}
\label{subsect:Ge(111)2x1}
%--------------------------------------------------------------------------%

%========================================================%
%========    Fig 2  " Cleaved_Ge2x1_surface "  ==========%
%========================================================%
% bb=0 0 998 1172
\begin{figure}
\includegraphics[width=83.5mm, scale=1.00]{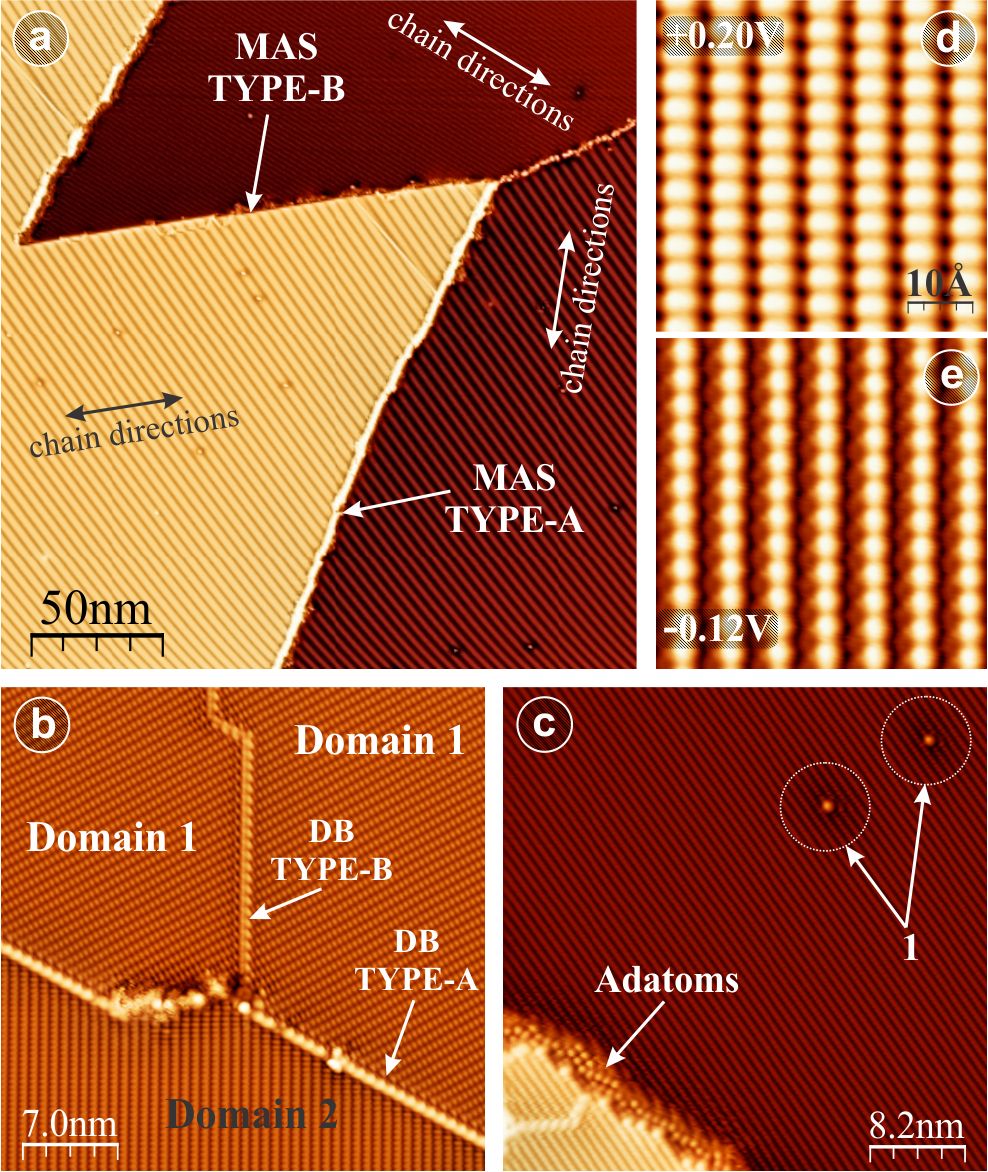}
\caption{(Color online) (a)-(c) Typical large scale STM topography images of the freshly cleaved Ge(111)2$\times$1 surface ($V_t = +1.0 \, {\rm
V}$, $I_{\rm t} = 35 \, {\rm pA}$). High resolution (d) empty and (e) filled states STM images of the same area, recorded at the indicated
tunneling voltage $V_{\rm t}$ and at $I_{\rm t} = 0.9 \, {\rm nA}$ and $3.0 \, {\rm nA}$, respectively.} \label{Cleaved_Ge2x1_surface}
\end{figure}
%========================================================%
%========================================================%
%========================================================%

The 2$\times$1 reconstruction of the Ge(111) surface consists of $\pi$-bonded chains of Ge atoms running in the $[01\overline{1}]$
direction.~\cite{Pandey_PRL_81, Northrup_PRB_83, Feenstra_SS_91} Only every other (upper chain) row can be imaged by STM.~\cite{Feenstra_PRB_01}
The surface unit cell contains two atoms, both having one dangling bond. This dangling bond is responsible for $\pi$-bonding along the upper
surface chain rows (see Fig.~\ref{2x1_schematic}). In the original Pandey geometry,~\cite{Pandey_PRL_81} the two upper atoms of the 7-member
ring have the same height and form ``zigzag'' chains along the $[01\overline{1}]$ direction. However, due to buckling, one of these two atoms
(\emph{up}-atom) is shifted upwards (out of the surface) while the other (\emph{down}-atom) is shifted downwards (into the surface), as
illustrated in Fig.~\ref{2x1_schematic}. The occupied surface states are mainly localized on the \emph{up}-atom, while the empty surface states
are mainly localized on the \emph{down}-atom. Consequently, the bonding surface states band $\pi_{\rm VB}$ derived from the \emph{up}-atom
orbital is filled, while the anti-bonding surface states band $\pi_{\rm CB}^*$ derived from the \emph{down}-atom orbital is empty.

In Fig.~\ref{Cleaved_Ge2x1_surface} we present typical large scale [(a)-(c)] and high resolution [(d)-(e)] STM topography images of the clean
Ge(111)2$\times$1 surface. Large atomically flat terraces up to $10^{5} \, {\rm nm}^{2}$ can be observed, which are separated from each other by
monatomic steps (MASs). It can be observed in Fig.~\ref{Cleaved_Ge2x1_surface}~(b) that the Ge(111)2$\times$1 surface consists of different
types of domains with slightly different atomic arrangement.~\cite{Einaga_PRB_98} This is related to the threefold rotational symmetry of the
surface. The domains are found to be separated by two different types of domain boundaries (DBs) [see Fig.~\ref{Cleaved_Ge2x1_surface}~(b)]. At
the first type of DB, referred to as a type-A DB following the terminology used in Ref.~[\onlinecite{Einaga_PRB_98}], the atomic rows at the
opposite sides of the DB are rotated by an angle $\pi/3$. The second type DB, the so-called anti-phase DB or type-B
DB,~[\onlinecite{Einaga_PRB_98}] is formed due to a shift of the $\pi$-bonded chain rows in the $[2\overline{11}]$ direction by half a unit
cell. We found that most DBs are of type-B and that the type-A DBs often exhibit (local) disorder.~\cite{Muzychenko_PRB_10}

%========================================================%
%========             Fig 3  " STS "           ==========%
%========================================================%
%, bb=0 0 1006 823
\begin{figure}
\includegraphics[width=85mm, scale=1.00]{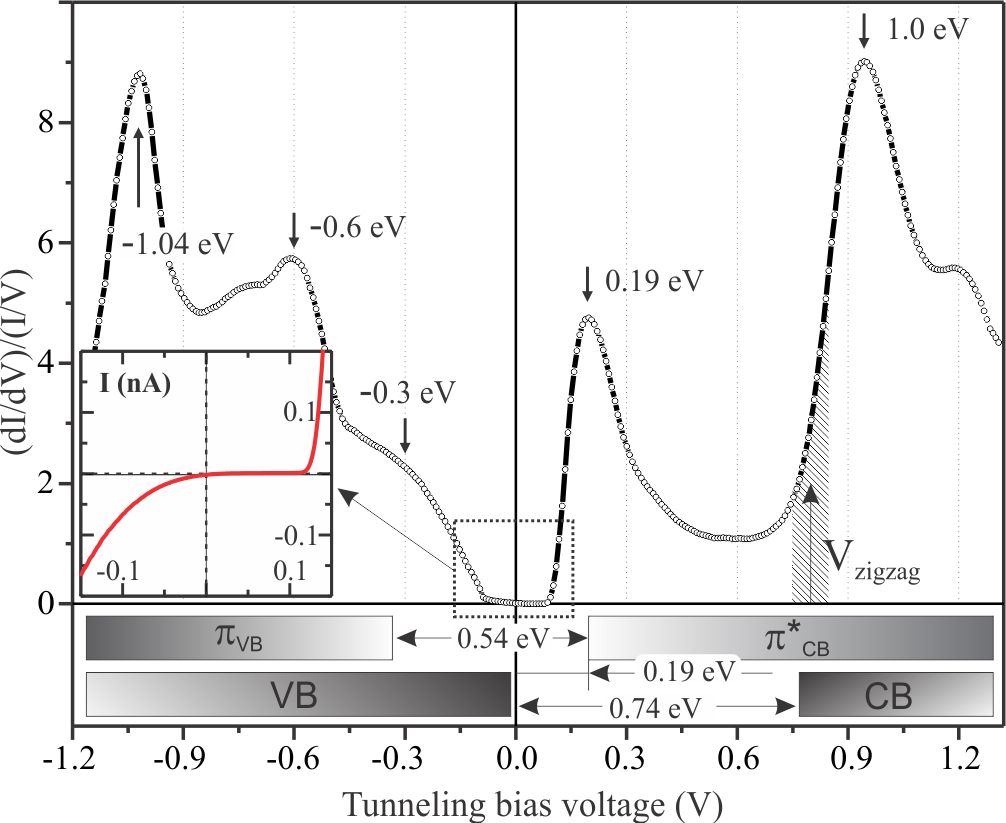}
\caption{Normalized conductance spectrum recorded on a defect free area of the freshly cleaved Ge(111)2$\times$1 surface. The main energy bands
are indicated by gray rectangles at the bottom of the figure (see text for more details). Inset: Current-voltage (\emph{I}-\emph{V})
characteristic close to the Fermi energy $E_{\rm F}.$} \label{STS}
\end{figure}
%========================================================%
%========================================================%
%========================================================%

In addition, two types of MASs can be observed [see Figs.~\ref{Cleaved_Ge2x1_surface}~(a) and (c)]. With respect to the upper terrace one type
of MASs, hereafter referred to as type-A, is oblique to the $\pi$-bonded chain rows on the terrace. The second type of MASs, hereafter referred
to as type-B, is parallel to the $\pi$-bonded chain rows. One should note that Fig.~\ref{Cleaved_Ge2x1_surface}~(a) exhibits pronounced Moir\'e
fringes that run along one direction and become visible because of the large size of the image,~\cite{Guo_Nanotech_04} while the $\pi$-bonded
chain rows are not visible on the STM image and their direction is specified by arrows for each of the terraces. We find that Ge adatoms are
often present at the MASs, both on the upper and on the lower terrace, except on the upper terrace of type-A MASs. At these terraces the surface
is either locally distorted or a 2$\times$4 or \emph{c}2$\times$8 surface reconstruction occurs
[Fig.~\ref{Cleaved_Ge2x1_surface}~(c)].~\cite{Feenstra_SS_91} The Ge surface adatoms are probably created upon cleavage at room temperature,
after which the adatoms can migrate along $\pi$-bonded chain rows to the MAS regions. Furthermore, (individual) Ge adatoms can be frequently
observed on atomically flat Ge(111)2$\times$1 terraces as well above a charged subsurface Ga impurity [see label (1) in
Fig.~\ref{Cleaved_Ge2x1_surface}~(c)]. These adatoms are well separated from each other and their number is in good agreement with the low
doping level of our Ge samples.

In Fig.~\ref{STS} we present a typical normalized conductance spectrum recorded at the \emph{p}-type Ge(111)2$\times$1 surface. The main energy
bands are indicated by gray rectangles at the bottom of Fig.~\ref{STS}. In the spectrum the large peak around $0.19 \ {\rm eV}$ can be assigned
to the onset of the unoccupied surface states conduction band $\pi_{\rm CB}^*$ related to the upper $\pi$-bonded chains of the Ge(111)2$\times$1
surface (see Fig.~\ref{2x1_schematic}).~\cite{Bechstedt_PRL_01, Kobayashi_PRB_03} Two energy gaps can be discerned: A narrow gap of about $0.19
\ {\rm eV}$ and a wide gap of about $0.74 \ {\rm eV}$. The latter corresponds to the forbidden energy gap of the projected bulk band structure
of the Ge(111) surface at low temperature. On the other hand, the narrow gap  corresponds to the energy gap between the filled bulk valence band
(VB) and the unoccupied surface states conduction band ($\pi_{CB}^*$). Using STS~\cite{Feenstra_PRB_01} as well as photoemission
experiments,~\cite{Nicholls_PRB_83, Nicholls_PRL_84, Nicholls_PRL_85} the surface states band gap has been determined before, yielding a gap
value of $0.54 \pm 0.04 \, {\rm eV}$. From Fig.~\ref{STS} it can be concluded that the high resolution STM images in
Fig.~\ref{Cleaved_Ge2x1_surface}~(e) and Fig.~\ref{Cleaved_Ge2x1_surface}~(d) were obtained at tunneling voltages near the top of VB and the
bottom of $\pi_{CB}^*$, respectively.

%--------------------------------------------------------------------------%
\subsection{Adsorption of Co atoms on Ge(111)2$\times$1}
\label{subsect:Co_adsorption}
\begin{figure}
\includegraphics[width=84.5mm, scale=1.00]{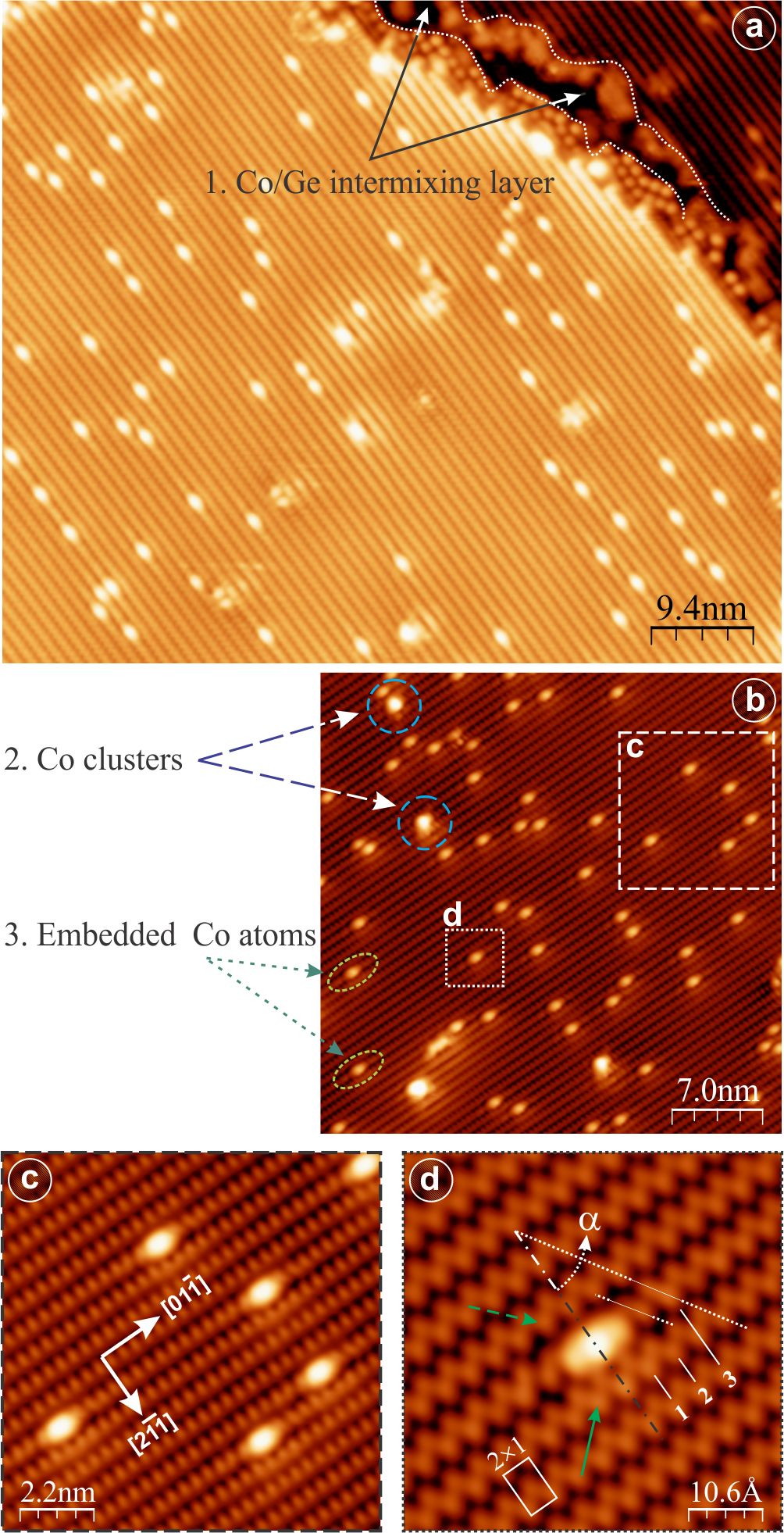}
\caption{(Color online) (a), (b) STM topography images of the Ge(111)2$\times$1 surface after deposition of $0.032 \, {\rm ML}$ of Co ($V_{\rm
t} = +1.0 \, {\rm V}$, $I_{\rm t} = 15 \, {\rm pA}$). (c), (d) High resolution STM images of the areas confined by the dashed rectangle and by
the dotted rectangle in (b), respectively ($V_{\rm t} = +0.9 \, {\rm V}$, $I_{\rm t} = 10 \, {\rm pA}$ for (c) and $V_{\rm t} = +0.80 \, {\rm
V}$, $I_{\rm t} = 300 \, {\rm pA}$ for (d)).} \label{Ge2x1+Co_atoms}
\end{figure}

%--------------------------------------------------------------------------%
In Figs.~\ref{Ge2x1+Co_atoms}~(a) and (b) we present two typical large scale STM topography images of the Ge(111)2$\times$1 surface after Co
atom deposition corresponding to a coverage of 0.032 ML. Three different kinds of structures are formed after Co deposition: (i) Co/Ge
intermixing layers (ILs) [indicated by the two arrows with label 1 in Fig.~\ref{Ge2x1+Co_atoms}~(a)], (ii) Co clusters consisting of multiple Co
atoms [indicated by the two arrows with label 2 in Fig.~\ref{Ge2x1+Co_atoms}~(b)] and (iii) well separated individual Co atoms [indicated by the
two arrows with label 3 in Fig.~\ref{Ge2x1+Co_atoms}~(b)].

The Co/Ge ILs are formed due to the consecutive accumulation of Co atoms at surface/subsurface defects, e.g. DBs and MASs.~\cite{Muzychenko_Submitted_11}
As indicated by the results of our DFT calculations that are presented below, Co atoms are able to migrate along the $\pi$-bonded chain rows, despite the
low temperature of the sample during Co deposition ($\rm T_{s} \leq 80 \, K$). Co/Ge ILs are found both on the upper and lower terraces at type-A MASs. At
the type-B MASs, formation of a Co/Ge IL occurs only on the lower terrace. Near DBs Co/Ge ILs are observed on both sides of the type-A and type-B DBs.
Co/Ge ILs can be found on atomically flat terraces as well, far away from any DBs and MASs. Since Co free atomic size defects, including in
particular Ga subsurface impurities,~\cite{Muzychenko_PRB_10} can no longer be observed after Co deposition, this suggest that these defects act as
nucleation centers for the formation of Co/Ge ILs. The amount of Co/Ge ILs formed on atomically flat terraces roughly scales with the amount of atomic
scale defects that is observed prior Co deposition. A more detailed discussion on the formation of Co/Ge ILs will be presented
elsewhere.~\cite{Muzychenko_Submitted_11}

%========================================================%
%=========    Fig 4    " Ge2x1+Co_atoms "     ===========%
%========================================================%
%, bb=0 0 1004 1962
%========================================================%
%========================================================%
%========================================================%

Only a small fraction of the Co atoms coagulates into small Co clusters. A larger fraction of the Co atoms remains under the form of individual
atoms after deposition. At tunneling voltages above $0.7 \, {\rm V}$ these individual atoms appear as bright protrusions located on the upper
$\pi$-bonded chain rows. At other tunneling voltages the Co atoms are observed differently. This voltage dependence is discussed in more detail
in Section~\ref{subsect:Co_bias_dep} below. High resolution STM topography images of individual Co atoms are presented in
Figs.~\ref{Ge2x1+Co_atoms}~(c) and (d). The amount of individual Co atoms that can be inferred from Figs.~\ref{Ge2x1+Co_atoms}~(a) and (b) is
$0.005 \pm 0.002 \, {\rm ML}$. $13 \pm 5$\% of the deposited amount of Co atoms is observed as individual Co atoms, while $87 \pm 5$\%
contributes to the formation of Co/Ge ILs and Co clusters. It is important to already note here that the individual Co atoms are actually not on
top of the Ge surface, as will be demonstrated in detail in Section~\ref{sect:Disc_and_DFT} below by comparing the experimental STM images to
simulated STM images based on DFT calculations. According to the DFT calculations, individual Co atoms penetrate into the Ge surface and reside
in between the $3^{\rm rd}$ and the $4^{\rm th}$ atomic layer (AL), in a quasi-stable position (at low temperatures) inside the 7-member Ge ring
of the 2$\times$1 reconstruction (see Fig.~\ref{2x1_schematic}). This ``embedding" is found to influence the local electronic structure, but
does not give rise to a modified surface reconstruction, as can be seen in Figs.~\ref{Ge2x1+Co_atoms}~(c) and (d). We therefore refer to this
embedding as ``noninvasive".

The low temperature of the Ge substrate during Co deposition appears to be crucial for obtaining individual, well separated Co atoms. As
mentioned above, in spite of the low substrate temperature, the larger fraction (around $87$\%) of the Co atoms still exhibits sufficient
surface mobility to migrate to defects where Co/Ge ILs are formed [see Fig.~\ref{Ge2x1+Co_atoms}~(a)]. The residual fraction of deposited Co
atoms (around $13$\%) remains confined to defect free atomically flat terraces of the 2$\times$1 surface. Recently, we reported that these Co
atoms diffuse from their quasi-stable sites to surface/subsurface defects as well after warming up the sample to room
temperature.~\cite{Muzychenko_Submitted_11} However, as long as the sample remains at low temperatures, the embedded individual Co atoms remain
immobile during the experiments in the investigated $-1.5 \, {\rm to} \, +1.5 \, {\rm V}$ voltage range.

As illustrated in Fig.~\ref{Ge2x1+Co_atoms}~(d), the ``zigzag"  structure of the 2$\times$1 reconstruction of the low doped \emph{p}-type
Ge(111) surface (see Fig.~\ref{2x1_schematic}) becomes observable for a limited range of tunneling voltages $V_{\rm zigzag} = 0.85\ \pm \ 0.07
\, {\rm V}$. From Fig.~\ref{STS} we can conclude that both the unoccupied surface states $\pi_{CB}^*$ (wave functions that are mainly localized
on the \emph{down}-atoms) and the unoccupied states at the bottom of conduction band (CB) (wave functions which are partially localized on the
\emph{up}-atoms~\cite{Nie_JVST_04}) become available for tunneling within this voltage range. Although the wave functions on the \emph{up}-atoms
have a smaller amplitude, the higher position of these atoms implies that they appear more prominently in the STM images than the
\emph{down}-atoms when the applied tunneling voltage increases. Within the $V_{\rm zigzag}$ voltage range (this range is marked by the gray
dashed area in Fig.~\ref{STS}), the contribution of $\pi_{CB}^*$ to the tunneling current remains nearly constant, while the contribution from
CB rapidly increases with increasing $V_{\rm t}$ up to $1.0 \, {\rm V}$. Hence, a balance exists between tunneling into \emph{up-} and
\emph{down-}atoms within the $V_{\rm zigzag}$ voltage range, which implies that both the \emph{up}- and \emph{down}-atoms of the $\pi$-bonded
chain rows are visualized in constant-current STM images [see Fig.~\ref{Ge2x1+Co_atoms}~(d)]. We found that the precise value of $V_{\rm
zigzag}$ depends on the doping level as well as on the semiconductor type. E.g., the 2$\times$1 reconstruction of the (111) surface of heavily
doped \emph{n}-type Ge(111) (phosphorus doping level $n_{\rm P} = 1 \times 10^{19} \, {\rm cm}^{-3}$) reveals zigzag chains around $0.55 \, {\rm
V}$. Previously, it has been reported by Trapmann \emph{et al.} that zigzag chains appear around $0.8 \, {\rm V}$ for \emph{n}-type
Si(111)2$\times$1 surfaces.~\cite{Trappmann_APA_99} The precise value of $V_{\rm zigzag}$ is therefore a characteristic feature of the
semiconductor surface under investigation.

%========================================================%
%========      Fig 5      " 5_ Co_atoms "      ==========%
%========================================================%
%, bb=0 0 922 768
\begin{figure}
\includegraphics[width=78mm, scale=1.00]{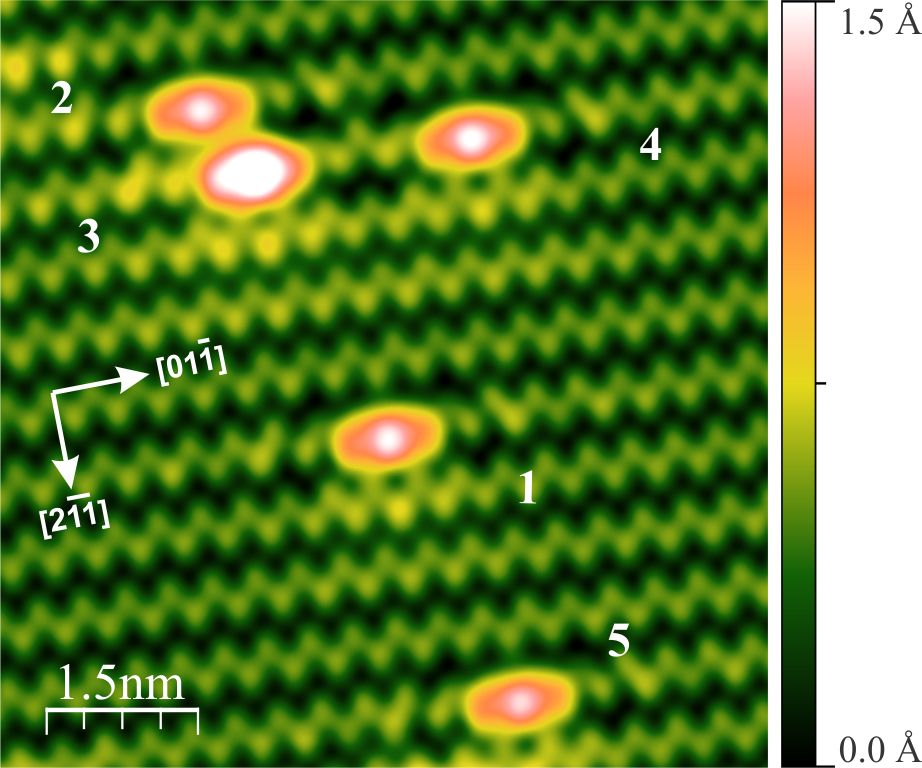}
\caption{(Color online) High resolution STM topography image of 5 Co atoms embedded in the Ge(111)2$\times$1 surface ($V_{\rm t} = +0.8 \, {\rm
V}$, $I_{\rm t} = 200 \, {\rm pA}$).} \label{5_Co_atoms}
\end{figure}
%========================================================%
%========================================================%
%========================================================%

From Figs.~\ref{Ge2x1+Co_atoms}~(a)-(d) it becomes clear that all individual Co atoms occupy identical positions at the Ge surface: Their
appearance is symmetric with respect to the $[2\overline{11}]$ direction [the symmetry axis is drawn as a black dash-dotted line in
Fig.~\ref{Ge2x1+Co_atoms}~(d)], while they appear asymmetric with respect to the $[01\overline{1}]$ direction. The latter can be related to the
asymmetry of the 2$\times$1 surface reconstruction (due to buckling) along this direction. The relevant crystallographic directions are
indicated in Fig.~\ref{Ge2x1+Co_atoms}~(c) and are the same for Figs.~\ref{Ge2x1+Co_atoms}~(b) and (d). The symmetry and asymmetry of the Co
atoms with respect to the $[2\overline{11}]$ and $[01\overline{1}]$ directions, respectively, becomes most clearly resolved for tunneling
voltages near $V_{\rm zigzag}$ [see Fig.~\ref{Ge2x1+Co_atoms}~(d) and Fig.~\ref{5_Co_atoms}]. The bright protrusion related to an individual Co
atom occupies about 2 surface unit cells along the $\pi$-bonded chain rows. Moreover, in Fig.~\ref{Ge2x1+Co_atoms}~(d), disturbance of the
zigzag atomic structure along the $[01\overline{1}]$ direction can be observed near the Co atom over a distance $\leq 3$ unit cell periods. In
Fig.~\ref{Ge2x1+Co_atoms}~(d) the edge of the unperturbed zigzag $\pi$-bonded chain on the right hand side of the Co atom is marked by a long
dotted white line, forming an angle $\alpha \approx 35 \, {\rm ^\circ}$ with the $[2\overline{11}]$ symmetry axis. Closer to the Co atom related
protrusion, a brighter/higher feature [marked by the short dotted white line and indicated by the label 2 in Fig.~\ref{Ge2x1+Co_atoms}~(d)] is
visible, also making an angle close to $\alpha$ with the $[2\overline{11}]$ symmetry axis. The same disturbance is observed to the left of the
Co atom (mirror symmetry with respect to the $[2\overline{11}]$ direction). Identical characteristic features are observed for all investigated
individual Co atoms on differently oriented Ge(111)2$\times$1 domains.

%========================================================%
%=======  Fig 6  " Multibias_Co_single_atoms "  =========%
%========================================================%
%, bb=0 0 2068 921
\begin{figure*}
\includegraphics[width=175mm, scale=1.00]{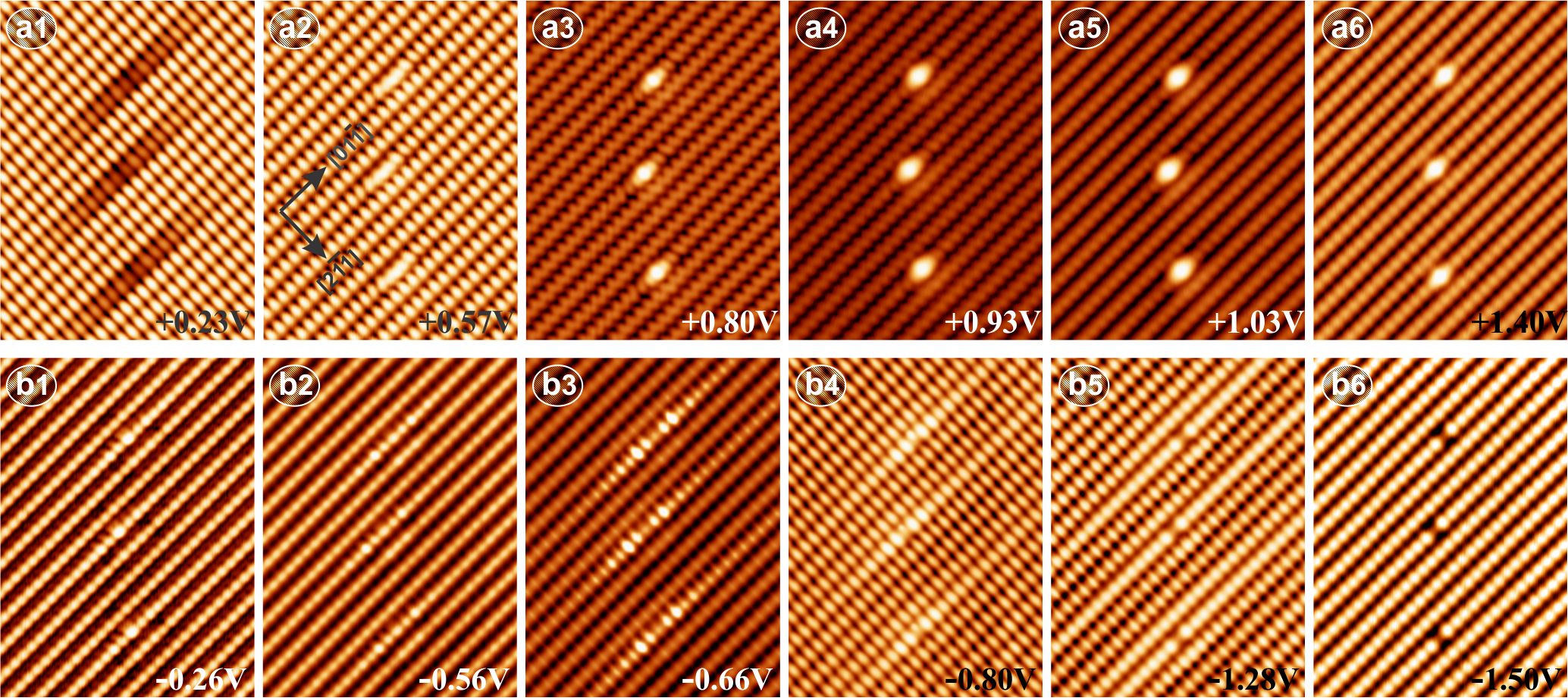}
\caption{(Color online) (a1)-(a6) Empty states  and (b1)-(b6) filled states STM topography images of 3 single Co atoms embedded in the
Ge(111)2$\times$1 surface. Image sizes are $15.0 \times 11.2 \, {\rm nm}^2$. The tunneling voltage $V_{t}$ is indicated for each image.
Tunneling current $I_t$ is $180 \,{\rm pA}$, $300 \,{\rm pA}$, $320 \,{\rm pA}$, $340 \,{\rm pA}$, $340 \,{\rm pA}$, and $340 \,{\rm pA}$ for
(a1,b1) to (a6,b6), respectively.} \label{Multibias_Co_single_atoms}
\end{figure*}
%========================================================%
%========================================================%
%========================================================%

In addition to the disturbance of the zigzag structure, an asymmetry of the electronic structure of the neighboring $\pi$-bonded chain rows
(with respect to the $[01\overline{1}]$ direction) near an embedded Co atom is observed. This asymmetry is most clearly resolved in maps of the
local density of states (LDOS) (not shown here). It can be observed as well in STM topography images around $V_{\rm zigzag}$: The lower
$\pi$-bonded chain row in Fig.~\ref{Ge2x1+Co_atoms}~(d) (indicated by the green solid arrow) appears brighter when compared to subsequent rows,
while the upper $\pi$-bonded chain row (indicated by the green dashed arrow) does not appear to be influenced by the embedded Co atom. This
asymmetric perturbation and the disturbance of the zigzag structure of the $\pi$-bonded chain rows always occur simultaneously and were observed
with various STM tips. Figure~\ref{5_Co_atoms} presents an STM topography image of 5 embedded single Co atoms that all induce a similar
difference between the Co neighbor chains. As we will demonstrate below when discussing the results of our DFT calculations, this additional
asymmetry of the Ge(111)2$\times$1 surface induced by the Co atoms allows us to determine the crystallographic $[2\overline{11}]$ direction of
the different Ge(111)2$\times$1 domains, leading to the conclusion that all investigated Ge(111)2$\times$1 surfaces consist of domains with
$\pi$-bonded chain-left isomers (negative buckling) exclusively.

There appears no significant electronic interaction between two embedded single Co atoms that reside in the same $\pi$-bonded chain row in close
vicinity (down to a distance of 3 unit cells along the $[01\overline{1}]$ direction) in the investigated voltage range from $-1.5 \, {\rm to} \,
+1.5 \, {\rm V}$ [see Figs.~\ref{Ge2x1+Co_atoms}~(a) and (b), as well as Co atoms (3) and (4) in Fig.~\ref{5_Co_atoms}]. On the other hand, an
electronic interaction effect can be observed for Co atoms that reside in neighboring $\pi$-bonded chain rows at a distance (in the $\pi$-bonded
chain row direction) smaller than $\pm$3 unit cells. This is illustrated in Fig.~\ref{5_Co_atoms}. Indeed, the lower Co atom (3), which is
located on a $\pi$-bonded chain row that is perturbed by the upper Co atom (2), exhibits modified electronic properties. More precisely, Co atom
(3) appears more bright in Fig.~\ref{5_Co_atoms}, while Co atom (2) appears similar to the other Co atoms. Note that the modified electronic
properties of Co atom (3) are caused by Co atom (2) only and are not related to presence of the neighboring Co atom (4) that is located in the
same $\pi$-bonded chain row.

%--------------------------------------------------------------------------%
\vspace{-2.0 mm}
\subsection{Voltage dependent STM investigation of single Co atoms embedded in Ge(111)2$\times$1}
\label{subsect:Co_bias_dep}
%--------------------------------------------------------------------------%
In this section we focus on the novel Co induced electronic features by careful comparison to the electronic properties of the clean
Ge(111)2$\times$1 surface (see Section~\ref{subsect:Ge(111)2x1}). First of all it must be noted that deposition of (a small amount of) Co atoms
does not change the electronic properties of the defect free Ge(111)2$\times$1 surface: The characteristic peaks in the STS spectra of the clean
Ge(111) surface prior to Co deposition (Fig.~\ref{STS}) are still observed at the 2$\times$1 reconstructed surface after Co deposition.
Figure~\ref{Multibias_Co_single_atoms} presents a series of (a) empty and (b) filled states STM topography images of 3 individual, well
separated Co atoms embedded in the reconstructed Ge(111)2$\times$1 surface. Images are recorded at the same location and with the same tip for a
broad range of tunneling voltages $V_{\rm t}$ between $1.4 \, {\rm V}$ and $-1.5 \, {\rm V}$. The series of images reveals a pronounced voltage
dependence for both the clean \emph{p}-Ge(111)2$\times$1 surface and the electronic influence of the embedded Co atoms on the $\pi$-bonded chain
rows.

In the empty state regime, at high voltages above $1.0 \, {\rm V}$ [Figs.~\ref{Multibias_Co_single_atoms}~(a5) and (a6)], topography is
dominated by the reconstruction lines of the $\pi$-bonded chain rows in the $[01\overline{1}]$ direction. Individual Co atoms appear as bright
protrusions, located directly on the upper $\pi$-bonded chain row and extending over 2 unit cells of the 2$\times$1 reconstruction. Maximum
contrast of the Co related features is observed around $V_{\rm t} = 0.93 \, {\rm V}$ and above this voltage the contrast again decreases. This
is because the contribution of the $\pi_{CB}^*$ surface states to the tunneling current remains approximately constant, while the contribution
of the bulk CB states increases with increasing tunneling voltage (see Fig.~\ref{STS}). At lower voltages around $V_{\rm zigzag} = 0.80 \, {\rm
V}$ [Fig.~\ref{Multibias_Co_single_atoms}~(a3)] the zigzag atomic structure discussed above emerges. Around $V_{\rm t} = 0.60 \, {\rm V}$
contrast of the Ge atomic corrugation and the Co related protrusions become similar and Co atoms can mainly be discerned by the locally induced
perturbation of the 2$\times$1 surface reconstruction [Fig.~\ref{Multibias_Co_single_atoms}~(a2)]. In the empty states regime below $0.60 \,
{\rm V}$ [Figs.~\ref{Multibias_Co_single_atoms}~(a1) and (a2)] extra corrugation appears in the STM topography along the $[2\overline{11}]$
direction. Close to the Fermi level $E_{\rm F}$ the strength of this extra corrugation along the $[2\overline{11}]$ direction becomes comparable
to the corrugation along the $[01\overline{1}]$ direction [Fig.~\ref{Multibias_Co_single_atoms}~(a1), also see
Fig.~\ref{Cleaved_Ge2x1_surface}~(d)]. Co atoms appear as centro-symmetric striped depressions along the $\pi$-bonded chain rows. These
depressions exhibit a local minimum directly above the embedded Co atom and gradually fade away with increasing distance (up to $4 \, {\rm nm}$)
from the local minimum [Fig.~\ref{Multibias_Co_single_atoms}~(a1)].

In the filled states regime, at voltages close to $E_{\rm F}$, topography is again dominated by corrugation of the $\pi$-bonded chain rows
[Fig.~\ref{Multibias_Co_single_atoms}~(b1), also see Fig.~\ref{Cleaved_Ge2x1_surface}~(e)]. Maxima of the atomic corrugation are related to the
highest filled bulk VB states that are localized on the Ge \emph{up}-atoms (see Fig.~\ref{2x1_schematic}). Co atoms appear as bright protrusions
located on the $\pi$-bonded chain rows [Fig.~\ref{Multibias_Co_single_atoms}~(b1)]. Below $-0.7 \, {\rm V}$ extra corrugation emerges along the
$[2\overline{11}]$ direction [Fig.~\ref{Multibias_Co_single_atoms}~(b4)] and persists down to around $-1.2 \, {\rm V}$. Below $-1.2 \, {\rm V}$
topography becomes again completely dominated by the $\pi$-bonded chain rows along the $[01\overline{1}]$ direction
[Fig.~\ref{Multibias_Co_single_atoms}~(b6)]. Here, Co atoms appear as atomic size vacancies in the upper $\pi$-bonded chain rows of the
2$\times$1 surface reconstruction [Figs.~\ref{Multibias_Co_single_atoms}~(b5) and (b6)].

It is clear from Figs.~\ref{Multibias_Co_single_atoms}~(a1) and (b2)-(b5) that the embedded Co atom induces a localized 1D perturbation of the
LDOS along the $\pi$-bonded chain rows. STM images recorded with a tunneling voltage near the edge of the surface states bands reveal highly
anisotropic scattering of electrons and screening effects with 1D confinement to the $\pi$-bonded chains. 1D perturbations have previously been
reported for Si(111)2$\times$1~\cite{Trappmann_APA_99, Garleff_PRB_05, Garleff_PRB_07} and Ge(111)2$\times$1~\cite{Muzychenko_PRB_10} surfaces.
Here, the perturbation near the atomic size defect has a pronounced 1D shape~\cite{Trappmann_APA_99} and extends up to $6 \, {\rm nm}$ along the
$\pi$-bonded chain rows, while the width of the perturbation remains limited to one period ($0.69 \, {\rm nm}$) of the 2$\times$1 reconstruction
in the $[2\overline{11}]$ direction. In the empty states regime, the 1D screening effects appear as depressions and become most pronounced near
$0.23 \, {\rm eV}$, corresponding to the bottom of the surface states band $\pi_{CB}^*$ [Fig.~\ref{Multibias_Co_single_atoms}~(a1)]. In the
filled states regime, the 1D electron scattering effects are observed as protrusions and are most clearly seen near $-0.66 \, {\rm V}$,
corresponding to the top of the surface states band $\pi_{VB}$ [Fig.~\ref{Multibias_Co_single_atoms}~(b3)]. In both regimes the 1D perturbations
exhibit identical mirror-like symmetry with respect to the $[2\overline{11}]$ direction, similar to our recent observations for atomic size
surface impurities on Ge(111).~\cite{Muzychenko_PRB_10}

In summary, we can state that individual Co atoms embedded in Ge(111)2$\times$1 surfaces exhibit the following general properties: \vspace{-1.5
mm}

\begin{itemize}
  \item Individual Co atoms penetrate into the cold ($\rm T_{s} \leq 80 \, K$) Ge(111)2$\times$1 surface after deposition (see Section~\ref{sect:Disc_and_DFT}
  below for more details); \vspace{-1.5 mm}

  \item the 2$\times$1 reconstruction is preserved after the noninvasive embedding of a Co atom; \vspace{-1.5 mm}

  \item the embedded Co atoms occupy identical positions in the Ge(111) surface and exhibit an identical voltage dependence of the STM topography
  images; \vspace{-1.5 mm}

  \item the embedded Co atoms exhibit a clear symmetry with respect to the $[2\overline{11}]$ direction, while they exhibit a clear asymmetry with
  respect to the $[01\overline{1}]$ direction; \vspace{-1.5 mm}

  \item the embedded Co atoms induce highly anisotropic scattering of electrons, which is accompanied by screening effects with 1D confinement
  along the $\pi$-bonded chain rows. \vspace{-1.5 mm}
\end{itemize}

%%%%%%%%%%%%%%%%%%%%%%%%%%%%%%%%%%%%%%%%%%%%%%%%%%%%%%%%%%%%%%%%%%%%%%%%%%%%
%%%%%%%%%%%%%%%%%%%%%%%%%%%%%%%%%%%%%%%%%%%%%%%%%%%%%%%%%%%%%%%%%%%%%%%%%%%%
%%%%%        Discussion and comparison with DFT calculations          %%%%%%
%%%%%%%%%%%%%%%%%%%%%%%%%%%%%%%%%%%%%%%%%%%%%%%%%%%%%%%%%%%%%%%%%%%%%%%%%%%%
%%%%%%%%%%%%%%%%%%%%%%%%%%%%%%%%%%%%%%%%%%%%%%%%%%%%%%%%%%%%%%%%%%%%%%%%%%%%
\section{Discussion and comparison with DFT calculations}
\label{sect:Disc_and_DFT}

%--------------------------------------------------------------------------%
\subsection{DFT model of the Ge(111)2$\times$1 surface}
\label{subsect:Ge_model}
\begin{figure}
\includegraphics[width=85.3mm, scale=1.00]{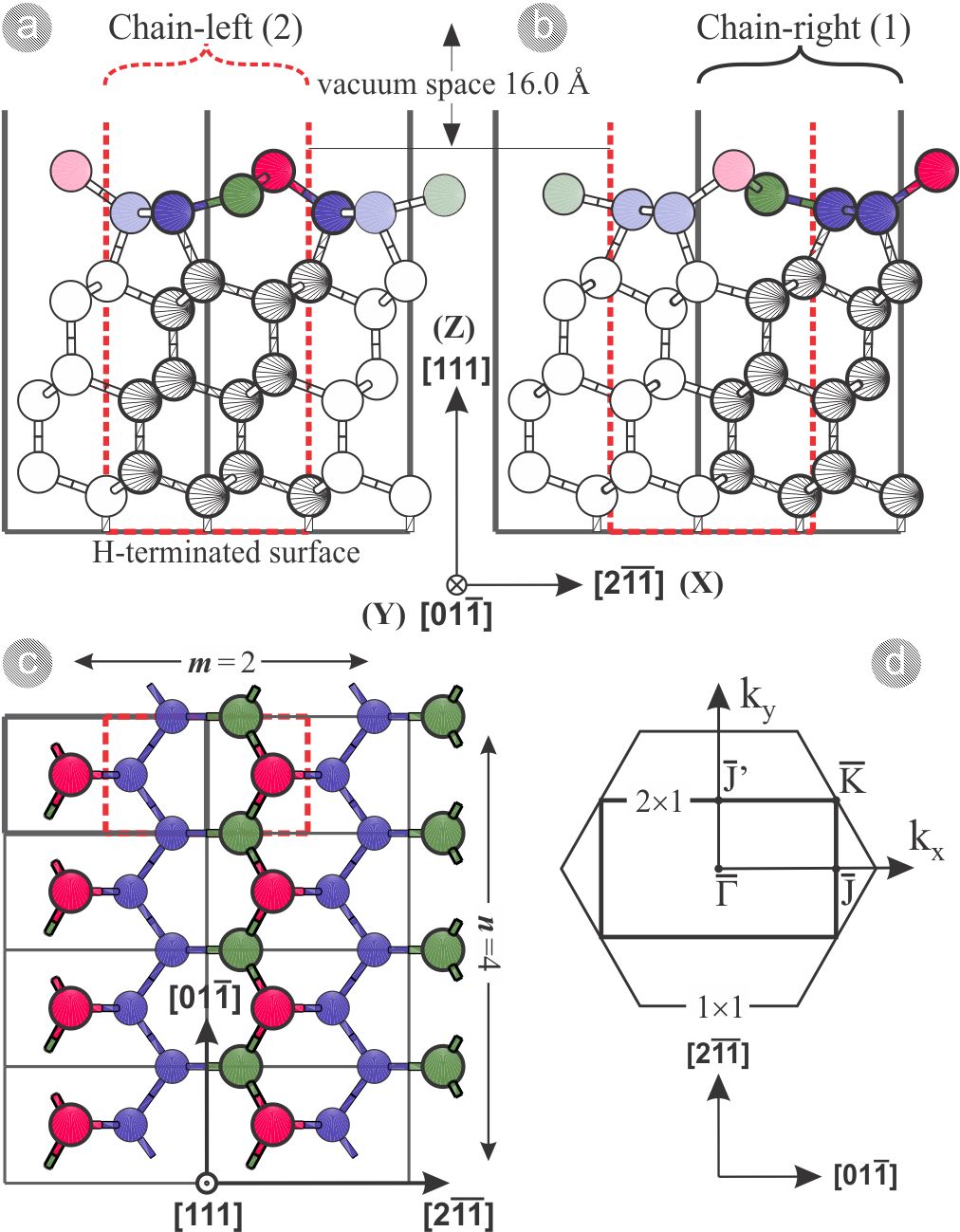}
\caption{(Color online) The final position of the 9 topmost Ge layers of the Ge(111)2$\times$1 surface reconstruction: (a) Chain-left and (b)
chain-right isomers of the Pandey $\pi$-bonded chain model with buckling (side view). (c) Top 3 surface atomic layers of the $4 \times 2$ SC
chain-left Ge(111)2$\times$1 isomer model (top view). (d) Surface Brillouin zone and relevant directions of the Ge(111)2$\times$1 surface.}
\label{Ge2x1_model}
\end{figure}

%--------------------------------------------------------------------------%
The 2$\times$1 reconstruction of cleaved Si and Ge (111) surfaces~\cite{Himpsel_PRB_84, Tromp_PRB_84, Alerhand_PRL_85, Northrup_PRL_91,
Feenstra_SS_91} is well described by the commonly used Pandey $\pi$-bonded chain model described above.~\cite{Pandey_PRL_81} The Pandey chain
geometry leads to a strong coupling of the dangling-bond orbitals along the chain, while the coupling between the chains is much weaker. This
geometry does not take into account the effects of buckling, so that the two uppermost surface Ge atoms in the 7-member rings are at the same
height.~\cite{Pandey_PRL_81} Northrup \emph{et al.} predicted buckling of the two uppermost atoms by about $0.8${\AA} (see
Fig.~\ref{2x1_schematic}).~\cite{Northrup_PRB_83} This buckling further reduces the surface energy of the system, yielding two different isomers
commonly referred to as $\pi$-bonded ``chain-left'' and ``chain-right'' isomers~\cite{Takeuchi_PRB_91} (see Fig.~\ref{Ge2x1_model}). Relying on
first principle calculations within the computational accuracy that could be achieved at that time (1991), it was found that the chain-left
isomer is $6 \, {\rm meV}$ per $1 \times 1$ surface cell lower in energy when compared to the chain-right isomer. It has been confirmed
theoretically~\cite{Rohlfing_PRL_00} and experimentally~\cite{Hirayama_PRB_00, Nie_JVST_04} that the chain-left isomer is indeed the dominant
isomer at the Ge(111)2$\times$1 surface. However, because of the very small energy difference between the two different isomers, coexistence of
both surface reconstructions and hence the possible existence of Ge(111)2$\times$1 multi-domain surfaces cannot be totally excluded.

Our theoretical investigation of the noninvasive embedding of Co atoms in the Ge(111)2$\times$1 surface was performed based on
DFT~\cite{Kohn_65} within the local density approximation (LDA).~\cite{Perdew_PRB_81} Calculations were performed with the SIESTA
package,~\cite{SIESTA_PRB_96, SIESTA_QC_97, SIESTA_CondMat_02} which relies on the expansion of the Kohn-Sham orbitals by linear combination of
pseudo-atomic orbitals. In all calculations a double-zeta basis set with polarization was used. The core electrons were implicitly treated by
using norm-conserving Trouiller-Martins pseudopotentials~\cite{TroullierMartins_91} with the following electronic configuration of the elements:
H $1s^1$, Ge (Ar $3d^{10}$) $4s^2 \, 4p^2$ and Co (Ar) $4s^2 \, 3d^7$, where the core configurations are indicated between parentheses. A cutoff
energy of $200 \, {\rm Ry}$ was introduced for the grid integration, ensuring convergence of the total energy of the system within typically
$0.1 \, {\rm meV}$.

%========================================================%
%==========    Fig 7     " Ge2x1_model "     ============%
%========================================================%
%, bb=0 0 1012 1299
%========================================================%
%========================================================%
%========================================================%

Our calculations are performed in three stages. First, we determine the surface equilibrium geometry for the chain-left and chain-right isomers within one
surface unit cell (SUC) using conjugate gradient (CG) geometry optimization. Second, based on the Ge(111)2$\times$1 SUC, a new $n \times m$ supercell (SC)
with one Co atom above the surface is constructed and the equilibrium configurations of Co/Ge(111)2$\times$1 are determined relying on CG geometry
optimization. Third, the electronic structure of a still larger SC is calculated, which is then used to perform the DFT based STM topography image
simulations.

In order to model the Ge(111)2$\times$1 surface, a reduced square SUC (unreconstructed) with ($a_0 \sqrt{2}, a_0 \sqrt{6}$) lattice vectors is
used ($a_0 = 2.8205$\AA \ is half the optimized bulk Ge lattice constant). This SUC consists of a slab of 26 Ge atomic layers, of which one
atomic layer is saturated by hydrogen atoms (52 Ge atoms and 2 H atoms per SUC). In the CG geometry optimization the 14 topmost atomic Ge layers
are allowed to move, while the 12 layers of Ge and H atoms are frozen at the ideal (bulk) positions. Using 130 \emph{\textbf{k}}-points within
the surface Brillouin zone, the atoms are relaxed until all atomic forces acting on the released atoms are smaller than $3 \, {\rm meV/}$\AA,
and the remaining numerical error in the total energy is smaller than $0.1 \, {\rm meV}$ for each optimization step.

The final positions of the 9 topmost Ge layers yielding the Ge(111)2$\times$1 surface reconstruction are presented in
Figs.~\ref{Ge2x1_model}~(a) and (c) (chain-left isomers) and Fig.~\ref{Ge2x1_model}~(b) (chain-right isomers). The coordinates of the Ge atoms
below the $7^{\rm th}$ atomic layer are found to change only slightly during the CG geometry optimization. For the considered SUC the two
different Ge isomers are both possible for the formation of the 2$\times$1 reconstruction. Depending on the initial conditions, either the
5-member or the 7-member Ge ring of the reconstruction is present in the 2$\times$1 SUC. This is illustrated in Figs.~\ref{Ge2x1_model}~(a) and
(c) for the chain-left isomer. The black solid bars in Figs.~\ref{Ge2x1_model}~(a) and (c) comprise the 5-member ring SUC, while the red dashed
bars comprise the 7-member ring SUC. The same can be done for the chain-right isomer, as illustrated in Fig.~\ref{Ge2x1_model}~(b). For the
5-member ring the upper $\pi$-bonded chain rows are formed at the joint of the SUC in the $[2\overline{11}]$ direction and are referred to as
``chain-left/right (1)'' hereafter [see, e.g., the solid colored atoms in Fig~\ref{Ge2x1_model}~(b)]. For the 7-member ring, the upper
$\pi$-bonded chain rows are formed inside the SUC and are referred to as ``chain-left/right (2)'' hereafter [see, e.g., the solid colored atoms
in Fig~\ref{Ge2x1_model}~(a)]. We verified that the chain-left/right (1) and chain-left/right (2) selection exhibit identical electronic
properties for periodical boundary conditions. Depending on the position of the adsorbed Co atom with respect to the upper $\pi$-bonded chain
rows in the $n \times m$ SC, selection (1) or (2) was chosen. Buckling distances of the chain-left and chain-right isomers are $0.83$\AA, and -
$0.80$\AA, respectively. The chain-right isomer has a total surface energy that is $14 \, {\rm meV/(2\times 1 \, SUC)}$ higher than the total
surface energy of the chain-left isomer, implying that the chain-left isomer should be the dominant isomer for the Ge(111)2$\times$1
surface.~\cite{Hirayama_PRB_00, Rohlfing_PRL_00, Nie_JVST_04}

%--------------------------------------------------------------------------%
\subsection{Co adsorption sites and energy decomposition}
\label{subsect:Co_sites}
%--------------------------------------------------------------------------%
To model the adsorption of a Co atom on the Ge(111)2$\times$1 surface, we  used the 9 topmost relaxed Ge layers for the chain-left and the
chain-right isomer that were obtained following the procedure described above (also see Fig.~\ref{Ge2x1_model}). 6 of these 9 atomic layers are
allowed to relax. The bottom side of the 3 fixed atomic layers is saturated by hydrogen atoms (16 Ge atoms and 2 H atoms per SUC; $16$ \AA \
slab vacuum space separation).

Geometry optimization was carried out for an enlarged 4$\times$2 SC for both the chain-left and the chain-right isomers. In
Fig.~\ref{Ge2x1_model}~(c) we show the three topmost layers of the 4$\times$2 SC (size is $15.95 \, {\rm \AA} \, \times \, 13.81 \, {\rm \AA}$,
consisting of 145 atoms) for the chain-left (1) isomer geometry. A Co atom was then located in front of the Ge(111)2$\times$1 surface. Next, the
relaxation of the Co atom was calculated by means of CG geometry optimization. CG geometry optimization for the 4$\times$2 SC was carried out
until all atomic forces acting on the released Co and Ge atoms were below $5 \ {\rm meV/ \AA}$ and until the numerical error on the total energy
was smaller than $10^{-4} \ {\rm eV}$ per SC for each optimization step. Using a variety of starting coordinates for the Co atom, multiple Co/Ge
quasi-stable geometries were tested and their total energies were compared. Both chain-left and chain-right isomer geometries were used. In both
cases multiple quasi-stable Co atom sites (with respect to an atomic force tolerance $3 \, {\rm meV/}$\AA) were found, including sites on the
Ge(111)2$\times$1 surface as well as underneath the Ge(111)2$\times$1 surface, i.e., inside the big 7-member Ge ring. The subsurface sites were
identified by using a location inside the 7-member ring as the starting location for the CG geometry optimization. On the other hand, the
surface sites were identified by using an initial Co location above the Ge(111)2$\times$1 surface, at heights in the 3 to $4 \, {\rm \AA}$ range
and at various initial coordinates in the ($x,y$)-plane. This way, 9 quasi-stable sites were identified for the chain-left isomer, of which 6
are located on the Ge surface and 3 are located inside the 7-member ring underneath the Ge surface. For the chain-right isomer 5 surface and 3
subsurface quasi-stable sites were identified.

%========================================================%
%========     Table 1     " Pos_energy "       ==========%
%========================================================%
\begin{table}[b]
\caption{Calculated difference $\triangle E_{\rm sites}$ in total energy for the Co atom located at the different sites (on top (1-6) and
underneath (7-9) the Ge(111)2$\times$1 surface) and the Co atom located at site (7). The calculated energy differences are for the chain-left isomer
geometry and are given for 4$\times$2, 8$\times$4 and 14$\times$4 SCs.} \label{Pos_energy}

\begin{ruledtabular}
\begin{tabular}{crrr}
\textrm{Co site number}& \textrm{4$\times$2 SC}& \textrm{8$\times$4 SC}&
\textrm{14$\times$4 SC}\\
\colrule
  (1) & $1.791\,{\rm eV}$ & moves to site (2) & - \\
  (2) & $0.376\,{\rm eV}$ & $0.724\,{\rm eV}$ & $1.250\,{\rm eV}$ \\
  (3) & $0.583\,{\rm eV}$ & $0.730\,{\rm eV}$ & $0.841\,{\rm eV}$ \\
  (4) & $0.966\,{\rm eV}$ & $1.199\,{\rm eV}$ & $1.350\,{\rm eV}$ \\
  (5) & $2.280\,{\rm eV}$ & $2.470\,{\rm eV}$ & - \\
  (6) & $0.701\,{\rm eV}$ & $1.054\,{\rm eV}$ & $1.280\,{\rm eV}$ \\
  (7) &     $0\,{\rm eV}$ &     $0\,{\rm eV}$ &     $0\,{\rm eV}$ \\
  (8) & $0.038\,{\rm eV}$ & $0.129\,{\rm eV}$ & $0.462\,{\rm eV}$ \\
  (9) & $0.512\,{\rm eV}$ & $0.595\,{\rm eV}$ & $0.693\,{\rm eV}$ \\
\end{tabular}
\end{ruledtabular}
\end{table}
%========================================================%
%========================================================%
%========================================================%

In Table~\ref{Pos_energy} we present for the chain-left isomer geometry and for all possible quasi-stable Co atom sites (labeled (1) to
(9) in the first column) an overview of the calculated difference $\triangle E_{\rm sites}$ in total energy with respect to the total energy for
the Co atom located at site (7). For a 4$\times$2 SC size (second column) the minimum energy was found to occur for the Co atom located at site
(7). Similar calculations were performed for the chain-right isomer geometry and for a
4$\times$2 SC (data not shown). Again, the minimum energy was found for the case of a Co atom located inside the 7-member ring, but with
somewhat lower energy gain when compared to the chain-left geometry.

In order to reduce the influence (related to the periodic boundary conditions) of the restructuring induced by the Co atom within a 4$\times$2
SC, similar calculations were performed for a 8$\times$4 SC (size is $31.90 \, {\rm \AA} \, \times \, 27.63 \, {\rm \AA}$, consisting of 577
atoms). The obtained values for the difference $\triangle E_{\rm sites}$ in total energy for all possible quasi-stable Co atom sites
after CG geometry optimization are listed in the third column of Table~\ref{Pos_energy}. Similar to the case of the 4$\times$2 SC, the energy
difference is again with respect to the total energy for the Co atom at site (7). When using this larger SC, a transition of the Co atom was
found from site (1) to site (2). Also, site (9) was found to be energetically more favorable than site (2) when compared to the 4$\times$2 SC.
This can be explained by the occurrence of longer-range surface relaxations inside the larger SC. Remarkably, for the 8$\times$4 and 14$\times$4
SCs all Co sites inside the 7-member ring have a lower energy (see Table~\ref{Pos_energy}) with respect to the other sites, which is different
from the 4$\times$2 SC.

%========================================================%
%==========    Fig 8     " Co_position "     ============%
%========================================================%
%, bb=0 0 1004 1188
\begin{figure}
\includegraphics[width=85mm, scale=1.00]{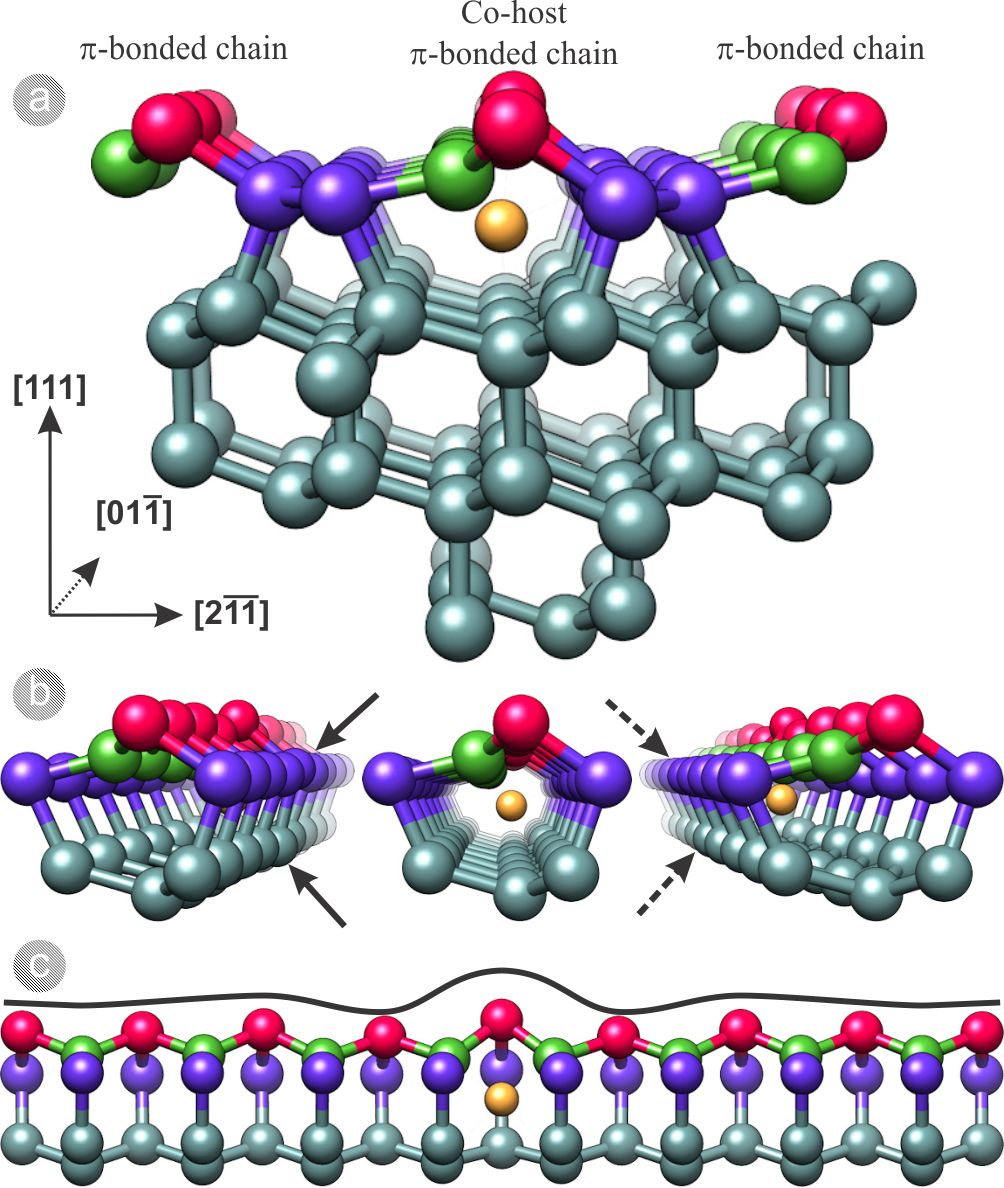}
\caption{(Color online) Calculated minimum energy position of a single Co atom (yellow) at the Ge(111)2$\times$1 surface. The Co atom is in
between the 3$^{\rm rd}$ and 4$^{\rm th}$ atomic layer underneath the Ge surface. (a), (b) 3D view of the chain-left Ge isomer. The Co atom is
located inside the 7-member Ge ring (viewed along the [01$\overline{1}$] direction). (c) Side view of the Co-host chain-left isomer (viewed
along the [2$\overline{11}$] direction). Height variations of the \emph{up}-atoms in the $\pi$-bonded chain row are indicated by the black
envelope curve (magnified by a factor 2 for clarity).} \label{Co_position}
\end{figure}
%========================================================%
%========================================================%
%========================================================%

Finally, calculated energy values for a  14$\times$4 SC (size is $55.82 \, {\rm \AA} \, \times \, 27.63 \, {\rm \AA}$, consisting of 1009 atoms)
are presented for the chain-left isomer geometry in the fourth column of Table~\ref{Pos_energy}. Similar calculations were performed for the
chain-right isomer geometry (data not shown) and yielded nearly identical results, with again a somewhat lower energy gain with respect to the
other sites when compared to the chain-left geometry.

%--------------------------------------------------------------------------%
\subsection{Prime location of the Co atom into Ge(111)2$\times$1 surface}
\label{subsect:Co_prime_loc}
%--------------------------------------------------------------------------%
In Fig.~\ref{Co_position}~(a) we present a 3D ball-and-stick model view of the relaxed Ge(111)2$\times$1 surface, obtained for the 8$\times$4 SC with
chain-left isomer geometry. The Co atom is located at the site of minimum energy, i.e. site (7) (see third column in Table~\ref{Pos_energy}). A 3D front view and a side view of the Co atom inside the 7-member ring are presented in
Fig.~\ref{Co_position}~(b) and in Fig.~\ref{Co_position}~(c), respectively. Relaxation of the surface Ge \emph{up}-atoms and \emph{down}-atoms upon Co atom
incorporation can be clearly observed. The black solid envelope line in Fig.~\ref{Co_position}~(c) reflects the variation of the \emph{z}-coordinate of the
center of the Ge \emph{up}-atoms (magnified by a factor 2 for clarity). The upward shift of the Ge \emph{up}-atom located directly above the Co atom is
$+0.43$\AA, while the downward shift of the neighboring Ge \emph{up}-atoms is $-0.10$\AA. The Co induced shift of the \emph{z}-coordinate of the Ge
\emph{up}- and \emph{down}-atoms extends as far as $\pm 3$ periods along the $\pi$-bonded chain row of the 2$\times$1 reconstruction, in agreement
with our experimental observations [see Fig.~\ref{Ge2x1+Co_atoms}~(d) and the related discussion in Section~\ref{subsect:Co_adsorption}].

The black solid arrows in Fig.~\ref{Co_position}~(b) (left image) indicate the bulk Ge atoms that experienced the most significant shift of their positions
upon embedding of the Co atom. In Fig.~\ref{Co_position}~(b) (right image) it can be seen that the Ge atoms to the left of the Co atom (indicated by the
black dashed arrows) remain unperturbed. This asymmetry of the geometry (and hence of the local electronic properties) of the Co/Ge(111)2$\times$1
system along the $[01\overline{1}]$ direction is in agreement with our experimental observations: The STM topography images in
Fig.~\ref{Ge2x1+Co_atoms}~(d) and Fig.~\ref{5_Co_atoms} also exhibit an asymmetry with respect to the $[01\overline{1}]$ direction around an embedded Co
atom. Finally, we want to stress once more that the embedding of Co atoms does not give rise to a novel Ge surface reconstruction. Instead, the Ge atoms
surrounding the Co atoms experience only small changes of their positions, which is accompanied by changes of the local electronic properties as well.
Experimentally, we also found that the 2$\times$1 reconstruction is maintained upon Co embedding, as becomes clear in
Figs.~\ref{Multibias_Co_single_atoms}~(a1), (a2), (b1) and (b2). Calculations for the other Co atom sites in Table~\ref{Pos_energy} reveal that these sites
lead to more drastic changes and in some cases even local destruction of the Ge(111)2$\times$1 reconstruction.

%--------------------------------------------------------------------------%
\subsection{Embedding of a Co atom into the Ge(111)2$\times$1 surface}
\label{subsect:Co_penetration}
%--------------------------------------------------------------------------%
In this section we will discuss the two most feasible routes for the incorporation/penetration of a Co atom, which is initially above the
Ge(111)2$\times$1 surface, to the site (7), i.e. inside the 7-member ring of the Ge(111)2$\times$1 surface. Embedding of deposited atoms into
subsurface layers has already been demonstrated before for a Si~\cite{Uberuaga_PRL_00} surface and for a Ge~\cite{Lin_PRB_92, Gurlu_PRB_2004}
surface. Ge atoms have been found both experimentally and theoretically to penetrate into the $4^{\rm th}$ subsurface layer of
Si(100)~\cite{Uberuaga_PRL_00} when deposited at a substrate temperature of about $500 {\rm ^\circ C}$. Similarly, it has been found that Si
atoms deposited on Ge(100)2$\times$1 are able to move below the Ge surface at room temperature.~\cite{Lin_PRB_92} Finally, the formation of
Co/Ge intermixing layers of up to 3 MLs thick after deposition of Co atoms on room temperature Ge(111) substrates has been demonstrated
experimentally.~\cite{Smith_JVSTA_89}

\begin{figure}
\includegraphics[width=84mm, scale=1.00]{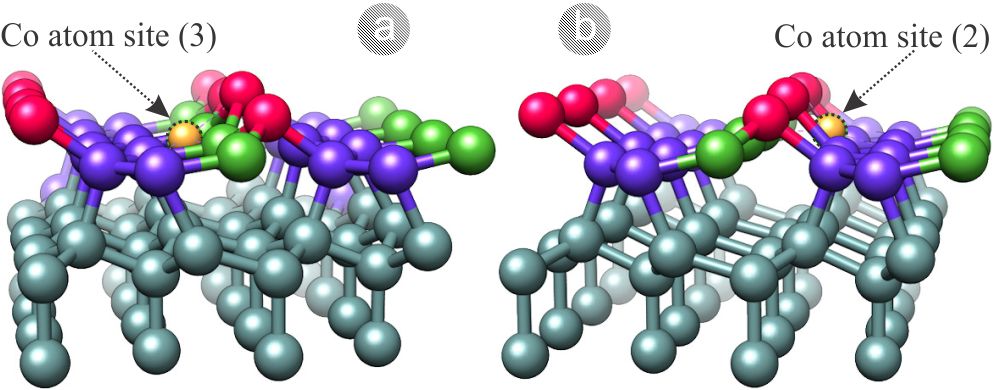}
\caption{(Color online) (a), (b) Calculated quasistable geometry ($4 \times 2$ SC) for Co sites (3) and (2) at the Ge(111)2$\times$1 surface
(see Table~\ref{Pos_energy}).} \label{Quasistable_23_pos}
\end{figure}

\begin{figure}
\includegraphics[width=84.8mm, scale=1.00]{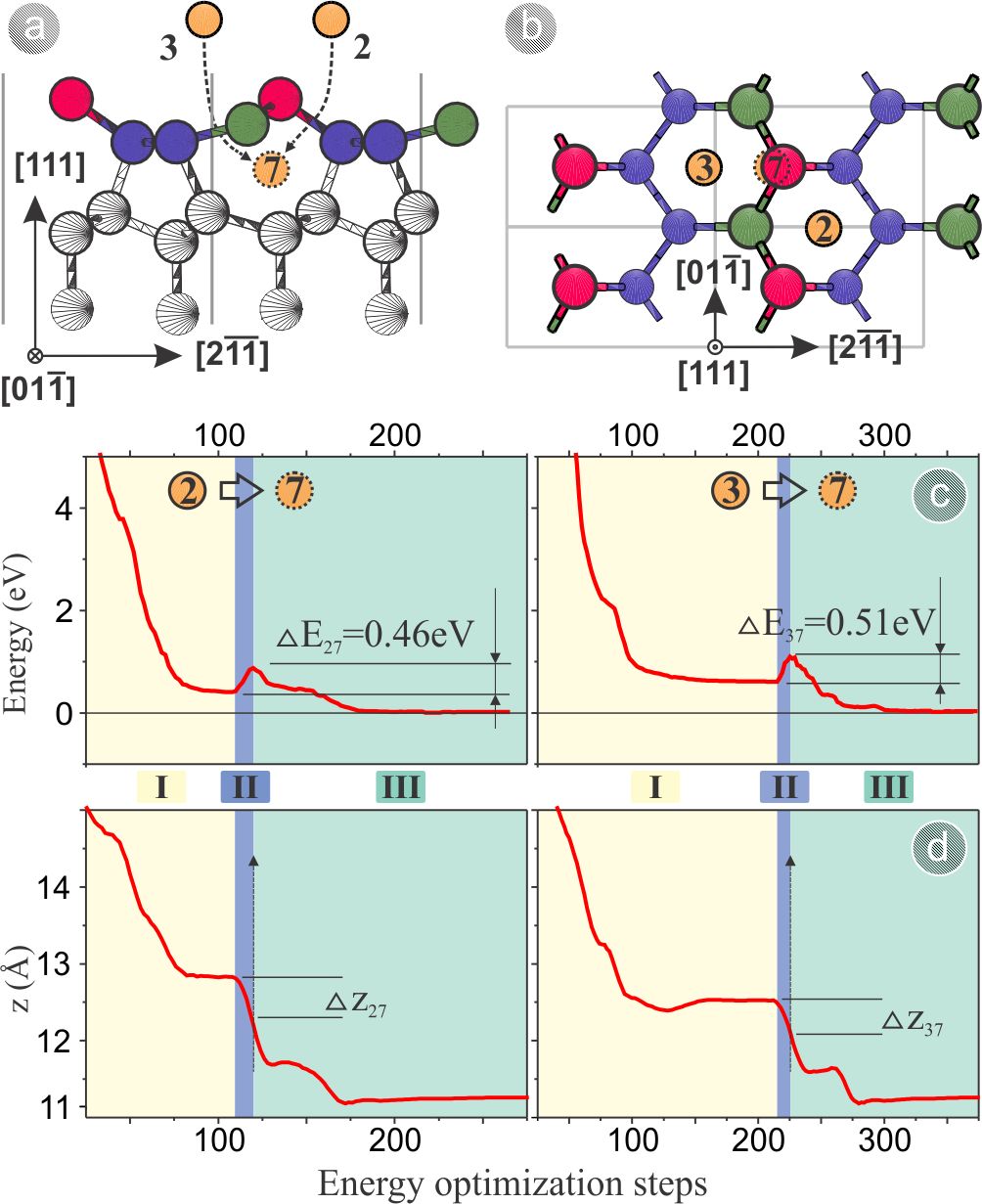}
\caption{(Color online) Two different routes are possible for the noninvasive embedding of a Co atom from the vacuum side of the Ge surface into
the big 7-member ring of the Ge(111)2$\times$1 surface reconstruction [also see Fig.~\ref{Co_position}~(a)]: From position (2) to (7) and from
position (3) to (7). (a) Side and (b) top view of the chain-left isomer of the Ge(111)2$\times$1 surface reconstruction. (c) Relative changes of
the total energy of the Ge(111)2$\times$1 4$\times$2 SC (including the Co atom) during subsequent geometry relaxation steps for both routes. (d)
Corresponding variation of the \emph{z}-coordinate of the Co atom.} \label{Co_penetration}
\end{figure}

As a possible starting site for penetration into the 7-member ring, the energetically two most favorable quasi-stable Co atom sites on top of
the Ge(111)2$\times$1 surface are considered: Site (2) and site (3) (see Table~\ref{Pos_energy}). A 3D visualization of both configurations is
presented in Fig.~\ref{Quasistable_23_pos}. The crystallographic directions of Fig.~\ref{Quasistable_23_pos} are identical to those of
Fig.~\ref{Co_position}~(a) (except for a rotation of the viewpoint angle with respect to the [111] direction). For the sites (2) and (3) the Co
atoms are located in between two neighboring upper $\pi$-bonded chain rows, at the center of the hexagonal 6-member Ge ring on the left
[Fig.~\ref{Quasistable_23_pos}~(a)] and right hand side [Fig.~\ref{Quasistable_23_pos}~(b)] of the central upper $\pi$-bonded chain row. A top
view of these sites at the Ge(111)2$\times$1 surface is presented in Fig.~\ref{Co_penetration}~(b). The Co atom at site (2) has one neighboring
Ge \emph{down}-atom and two neighboring Ge \emph{up}-atoms, while the Co atom at site (3) has one neighboring Ge \emph{up}-atom and two
neighboring Ge \emph{down}-atoms. One should note that for both sites the Co atom is already somewhat below the $1^{\rm st}$ Ge layer, having a
\emph{z}-coordinate comparable to that of the \emph{down}-atom for site (2) and $0.1 \, \rm \AA$ lower than the \emph{down}-atom for the site
(3). As was the case for site (7), these Co atom sites do not destroy the reconstruction of the Ge(111)2$\times$1 surface.

%========================================================%
%========   Fig 9    " Quasistable 23_pos "    ==========%
%========================================================%
%, bb=0 0 994 390
%========================================================%
%========================================================%
%========================================================%

Next, both quasi-stable Co positions were taken as the starting positions for additional first principles DFT calculations with a 4$\times$2 SC
and using the same parameters as described above. The Co atom was forced to ``move" into the bulk of the Ge by sequential decrements of its
\emph{z}-coordinate ($\triangle z = 0.04 \, \rm \AA$ for each geometry relaxation step). After each forced sequential decrement of the
\emph{z}-coordinate, the position of the Co atom is kept fixed, while the Ge atoms are allowed to relax and the total energy is determined. By
monitoring the total energy of the 4$\times$2 SC during subsequent geometry relaxation steps, we are able to evaluate the potential barrier
height that needs to be overcome by the Co atom when diffusing either from site (2) or from site (3) to site (7) without destroying the
2$\times$1 surface reconstruction. First, when the Co atom is forced to move only very slightly below the Ge surface, the Co atom ``bounces"
back to its initial position. Second, upon a certain minimum translation $\triangle z$ along the \emph{z}-direction (corresponding to a
potential barrier $\triangle E$), the released Co atom continues to move further below the Ge surface to site (7) [see
Figs.~\ref{Co_penetration}~(a) and (b)]. Both Co atom trajectories are visualized schematically in Fig.~\ref{Co_penetration}~(a).

%========================================================%
%========     Fig 10   " Co_penetration "      ==========%
%========================================================%
%, bb=0 0 1004 1231
%========================================================%
%========================================================%
%========================================================%

The variation of the total energy $E_{\rm tot}$ of the 4$\times$2 SC (again with respect to $E_{\rm tot}$ of the 4$\times$2 SC with the Co atom at site
(7)) and the variation of the \emph{z}-coordinate of the Co atom during the subsequent geometry relaxation steps are presented in
Fig.~\ref{Co_penetration}~(c) and in Fig.~\ref{Co_penetration}~(d) for the two considered routes. Three regimes can be discerned. Regime I describes the
energy gain and \emph{z}-coordinate variations during movement of the Co atom from vacuum to either site (2) or site (3). Regime II corresponds to the
``forced movement" of the Co atom into the bulk of the Ge up to the ``point of no return" when the potential barrier has been overcome. Finally, regime III
reflects the subsequent relaxation of the Co atom towards its final and stable position at site (7) in the 7-member Ge ring. The number of geometry
relaxation steps for route (2)$\Rightarrow$(7) and for route (3)$\Rightarrow$(7) is 500 and 700, respectively. For clarity, geometry relaxation intervals
where the total energy $E_{\rm tot}$ and the $z$-value remain almost constant (at the end of the regimes I and III) are cut from
Figs.~\ref{Co_penetration}~(c) and (d). The graphs presented in Fig.~\ref{Co_penetration}~(c) allow us to determine the potential barrier that the Co atom
needs to overcome for route (2)$\Rightarrow$(7) [(3)$\Rightarrow$(7)]: $\triangle E_{\rm 27} = 0.46 \, {\rm eV}$ [$\triangle E_{\rm 37} = 0.51 \, {\rm
V}$], corresponding to a change in \emph{z}-coordinate $\triangle z_{\rm 27} = 0.48 \, \rm \AA$ [$\triangle z_{\rm 37} = 0.40 \, \rm \AA$] in
Fig.~\ref{Co_penetration}~(d).

To overcome the surface potential barrier for penetration below the Ge surface, a Co atom must have a sufficiently high (kinetic) energy upon
deposition. In our experiments Co atoms are evaporated using an e-beam evaporator, where the Co material is heated to a high temperature
$T_{\rm v}$ around $3000 \, {\rm K}$.~\cite{Asano_JNST_92, Miller_88} Atoms leaving an e-beam melt generally have a narrow energy
distribution~\cite{Chopra_69} and the kinetic energy of the evaporated cloud of Co atoms, which is induced by the high temperature evaporation
process can be roughly estimated using the equipartition theorem, yielding a mean atomic kinetic energy of about $0.38 \, {\rm eV}$. Recently,
however, Asano \emph{et al.} have demonstrated experimentally that the velocity of evaporated atoms is typically even higher than the maximum
velocity suggested by the ideal gas approximation.~\cite{Asano_JNST_92} The increased velocity can be accounted for by a conversion of electron
excitation energy to kinetic energy during the adiabatic expansion away from the heated material. If the gas flow of evaporated atoms cooled
sufficiently during the adiabatic expansion, the resulting maximum velocity can be estimated as~\cite{Uetake_RSI_91}

\begin{equation}
  v_{\rm max} = \sqrt{\frac{2}{m}\cdot \frac{\gamma}{\gamma - 1} R \, T_{\rm v}},
\label{Atom_vel}
\end{equation}

\noindent where $\gamma$ is the specific heat ratio $C_{\rm p}/C_{\rm v}$ ($C_{\rm p}$ and $C_{\rm v}$ are the specific heat at constant
pressure and at constant volume per mole, respectively), $R$ is the gas constant and $m$ is the molar mass of the evaporated atom. $\gamma$ is
$5/3$ for an ideal mono-atomic gas. For $T_{\rm v} \simeq 3000 \, {\rm K}$, Eq.~(\ref{Atom_vel}) yields a maximum velocity $v_{\rm max} = 1460
\, {\rm m/s}$ and hence the kinetic energy of the deposited Co atoms may exceed even a maximum value of $0.63 \, {\rm eV}$. This kinetic energy
allows a Co atom to overcome the surface energy barrier $\triangle E_{\rm 27} = 0.46 \ {\rm eV}$ or $\triangle E_{\rm 37} = 0.51 \ {\rm eV}$
that is encountered when penetrating below the Ge(111)2$\times$1 surface following the route (2)$\Rightarrow$(7) or the route
(3)$\Rightarrow$(7), respectively [see Fig.~\ref{Co_penetration}~(c)].

%--------------------------------------------------------------------------%
\subsection{DFT-based modeling of STM topography images}
\label{subsect:STM_modeling}
%--------------------------------------------------------------------------%

%========================================================%
%========   Fig 11   " STM_DFT_comparision "   ==========%
%========================================================%
%, bb=0 0 2046 1651
\begin{figure*}
\includegraphics[width=172.5mm, scale=1.00]{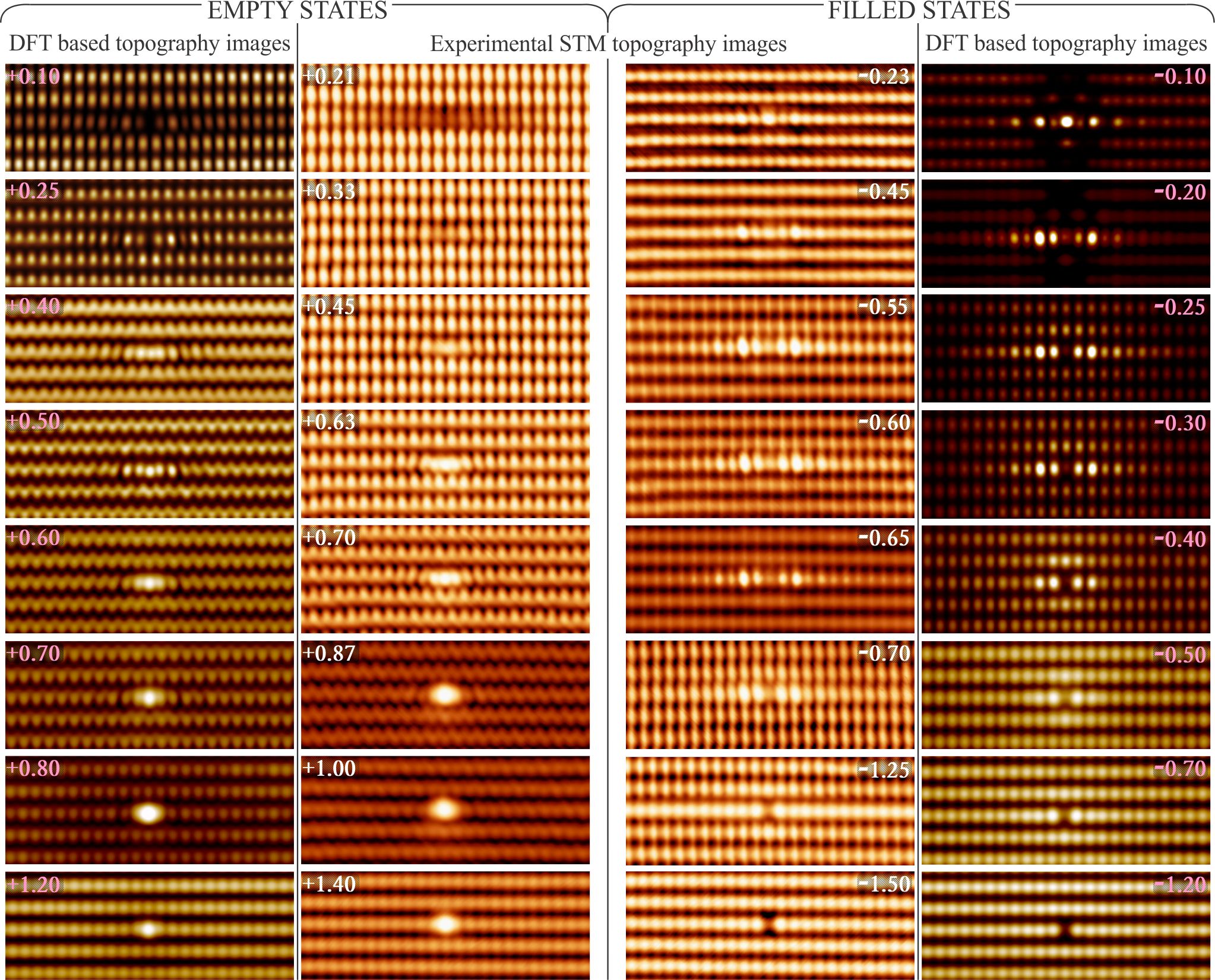}
\caption{(Color online) $9.2 \times 3.5 \, {\rm nm}^2$ experimental empty and filled states STM topography images (inner columns), together with
the corresponding calculated DFT-based STM topography images (outer columns) of a single Co atom located in a subsurface 7-member Ge ring, in
between the $3^{\rm rd}$ and $4^{\rm th}$ atomic layer underneath the Ge(111)2$\times$1 surface. The tunneling voltage $V_t$ is indicated for
each image.} \label{STM_DFT_comparision}
\end{figure*}
%========================================================%
%========================================================%
%========================================================%

Our theoretical findings, related to the changes in surface energy and the potential barrier for penetration below the surface, support the idea
of a noninvasive embedding of individual Co atoms in subsurface 7-member Ge rings. To verify the proposed ``embedding model", we investigated
the electronic properties of the Co/Ge(111)2$\times$1 system by simulating the corresponding STM topography images using DFT based calculations
for a wide range of voltages, which allows for a direct and detailed comparison between theory and experiment. For this purpose we investigated
the electronic structure of the Ge(111)2$\times$1 surface for each of the possible Co atom locations described in
Section~\ref{subsect:Co_sites}, for both chain-left and chain-right isomers. The calculated quasi-stable geometries of the relaxed 8$\times$4
SCs (9 and 8 possible geometries for the chain-left and chain-right isomer, respectively) were transferred to a larger 14$\times$4 SC (size is
$55.82 \, {\rm \AA} \, \times \, 27.63 \, {\rm \AA}$) for which we calculated the electronic properties in detail.

In order to construct STM constant current topography images based on the calculated electronic structure of the Co/Ge(111)2$\times$1 system,
the decay of the electron wave functions from the surface into the vacuum needs to be taken into account. Within the Tersoff-Hamann
theory~\cite{Tersoff_PRL_83, Tersoff_PRB_85} the dependence of the tunneling current $I$ on the applied tunneling voltage $V_{\rm t}$ between an
STM tip and a surface is given by

\begin{equation}
  I = \frac{2 \pi e}{\hbar} \sum_{\mu,v} f(E_{\mu})[1-f(E_v+eV_{\rm t}]\mid M_{\mu v}\mid ^2 \delta(E_{\mu} -E_v),
\label{Tunneling_curent}
\end{equation}

\noindent where $f(E)$ is the Fermi function, $M_{\mu v}$ is the tunneling matrix element between electronic states $\psi_{\mu}$ of the tip and
electronic states $\psi_{v}$ of the surface, and $E_{\mu}$ ($E_{v}$) is the energy of the state $\psi_{\mu}$ ($\psi_{v}$) in the absence of
tunneling. When we assume localized wave functions $\psi_{\mu}$ for the tip, $M_{\mu v}$ will vary proportional to the amplitude of $\psi_v$ at
the position $\overrightarrow{r}_0$, which corresponds to the center of the sphere that is used to approximate the tip apex. At low temperatures
and for small tunneling voltages $V_{\rm t}$ Eq.~(\ref{Tunneling_curent}) reduces to

\begin{equation}
  I \propto \sum_{v} \mid \psi_{v}(\overrightarrow{r}_0)\mid ^2 \delta(E_{\mu} -E_F).
\label{Tunneling_curent2}
\end{equation}

From Eq.~(\ref{Tunneling_curent2}) it follows that the tunneling current $I$ is proportional to the surface LDOS that is probed at position
$\overrightarrow{r}_0$ of the tip, integrated over an energy range from $E_{\rm F}$ to $E_{\rm F} + e V_{\rm t}$. For constant tunneling current
$I = I_{t}$ the STM tip essentially follows a contour of constant surface LDOS. However, because the surface wavefunctions decay exponentially
into the vacuum region, numerical evaluation of $\psi_v(\overrightarrow{r}_0)$ for tip-surface distances of the order of several angstroms
becomes a significant problem for DFT calculations.~\cite{Paz_PRL_05} For this reason STM simulations are often restricted to (the vicinity of)
the surface, which may yield incorrect results. To tackle this problem, we have used the 2D Fourier transform of the wavefunctions
$\psi_v(\overrightarrow{r})$ in combination with spatial extrapolation techniques~\cite{Rohlfing_PRB_07} to evaluate the surface wave function
$\psi_v(x,y,z)$ in the vacuum region up to $z = 7 \, {\rm \AA}$ above the surface.

This way we calculated the STM topography images for all available Co atom sites and for both the chain-left and chain-right isomers (for a 14$\times$4 SC)
within an energy range between $-1.5 \, {\rm eV}$ and $+1.5 \, {\rm eV}$ and at distances up to $7 \, {\rm \AA}$ above the surface. Perfect agreement
between theory and experiment for the whole energy range can be achieved only for a Co atom located at site (7) for the chain-left isomer geometry. For site (7) we then calculated the electronic properties also for a 26$\times$4 SC (size is $103.67 \, {\rm \AA} \, \times \, 27.63 \, {\rm \AA}$, consisting of 1873 atoms) and a 23$\times$5 SC (size is $91.74 \, {\rm \AA} \, \times \, 34.54 \, {\rm \AA}$, consisting
of 2071 atoms). CG geometry optimization for the chain-left(2) isomer with the Co atom located at site (7) within a 9$\times$3
($35.89$~\AA~$\times$~$20.72$~\AA) SC was performed, similar to the calculations described in Section~\ref{subsect:Co_sites}. We used the chain-left (1)
and chain-left (2) isomers (see Section~\ref{subsect:Ge_model}) for the 26$\times$4 SC and 23$\times$5 SC, respectively, to keep the Co atom in the center
of the SC. Calculations for the 26$\times$4 chain-left (1) SC and the 23$\times$5 chain-left (2) SC yield identical results.

For our simulations of the STM topography images we have to rely on experimental $z(V_{\rm t})$ spectra measured on the Ge(111)2$\times$1
surface in order to take into account the dependence of the height \emph{z} on the tunneling voltage $V_{\rm t}$ in our calculations. The
experimental $z(V_{\rm t})$ dependence with an initial height addition of $3 \, {\rm \AA}$~\cite{Feenstra_PRB_91} was used to determine the
height above the Ge(111)2$\times$1 surface at which simulated STM images are calculated. For low voltages, i.e. for energies close to $E_{F}$,
$z(V_{\rm t}) \simeq 3.7 \, {\rm \AA}$, while for high voltages above $1 \, {\rm V}$, $z(V_{\rm t}) \simeq 6 \, {\rm \AA}$.

In Fig.~\ref{STM_DFT_comparision} we present a series of experimental (inner columns) and calculated (outer columns) STM topography images for
the filled (two right columns) and empty (two left columns) states regime within a wide range of tunneling voltages. Calculated STM topography
images are obtained for a 23$\times$5 SC with chain-left (2) isomer geometry with the Co atom at site (7). Experimental STM topography images
are all recorded at the same location. The Co atom is well separated from other Co atoms, imlying there is no influence from neighboring Co
atoms (see Section~\ref{sect:Exp_Res}).

As can be seen in Fig.~\ref{STM_DFT_comparision}, correspondence between the Co related features in the calculated and the experimental STM
topography images is striking for the whole investigated voltage range. Concerning the precise tunneling voltage $V_{\rm t}$ at which optimum
correspondence is observed between theory and experiment there is a minor mismatch. This mismatch exhibits a non-linear dependence on the
applied tunneling voltage for both the filled and empty states regime. At low tunneling voltages the difference in voltage is around $0.1 \ {\rm
V}$ for both the empty and filled states regime (see the first row of images in Fig.~\ref{STM_DFT_comparision}). For higher tunneling voltages,
the difference increases to around $0.37 \ {\rm V}$ and $0.23 \ {\rm V}$ for the filled and empty states regime, respectively. Upon more careful
comparison, it can be seen that the difference in voltage mainly affects the surrounding Ge(111)2$\times$1 surface and not the Co atom itself.
Indeed, maximum contrast related to the Co protrusion appears around $V_{\rm t} = 0.9 \pm 0.1 \ {\rm V}$ in both the experimental
[Fig.~\ref{STM_DFT_comparision} and Fig.~\ref{Multibias_Co_single_atoms}~(a4)] and the calculated STM topography images. Apart from the rather
small difference in voltage, there is a very good agreement between the theoretically and experimentally observed electronic features for both
the filled and empty states regime.

The results presented in Fig.~\ref{STM_DFT_comparision} confirm that the calculated images of the Co/Ge(111) system exhibit all typical features
that were observed in the voltage dependent STM investigation (Section~\ref{subsect:Co_bias_dep}): (i) The Co induced strongly perturbed area
comprises $\pm$2 SUCs on the upper $\pi$-bonded chain row at moderate and high energies in the empty states regime; (ii) the perturbation has a
mirror symmetry axis along the $[2\overline{11}]$ direction; (iii) the perturbation exhibits a clear asymmetry with respect to the
$[01\overline{1}]$ direction; (iv) the Ge(111)2$\times$1 surface exhibits a zigzag structure only at a specific tunneling voltage and this
zigzag structure is perturbed near the Co atom.

In the filled states regime the calculated 1D Co induced perturbation along the $\pi$-bonded chain row exhibits the highest contrast in the
$-0.25 \ {\rm eV} \, {\rm to} \, -0.35 \ {\rm eV}$ voltage range, whereas experimentally the highest contrast occurs around $e V_{\rm t} = -0.65
\pm 0.05 \ {\rm eV}$ [see Fig.~\ref{Multibias_Co_single_atoms}~(b3) and Fig.~\ref{STM_DFT_comparision}]. At these energies the Co induced
perturbed atomic corrugation along the upper $\pi$-bonded chain row in the calculated and experimental STM topography image match very well. At
energies below $-1.0 \ {\rm eV}$ the Co atom appears as an atomic size vacancy in the upper $\pi$-bonded chain row for both the calculated and
experimental STM topography.

%========================================================%
%========     Fig 12    " Calc Co Zigzag  "    ==========%
%========================================================%
%, bb=0 0 997 374
\begin{figure}
\includegraphics[width=84.2mm, scale=1.00]{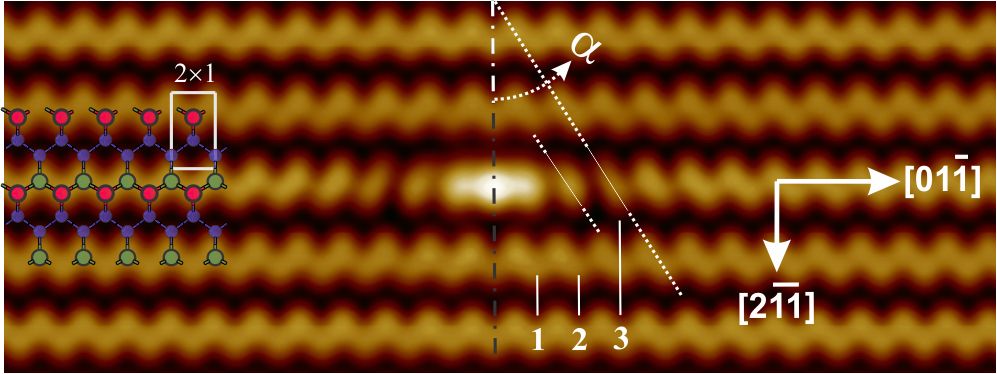}
\caption{(Color online) Calculated empty state STM topography image of a Co atom located at site (7) in the Ge(111)2$\times$1 surface. The empty
state energy ($0.57 \, {\rm V}$)is close to the energy for optimum calculated contrast of the zigzag structure of the upper $\pi$-bonded
chains.} \label{Calc_Co_zigzag}
\end{figure}
%========================================================%
%========================================================%
%========================================================%

In the calculated images the zigzag structure of the Co containing upper $\pi$-bonded chain rows is observed at empty state energy
around $0.57 \, {\rm eV}$ (see Fig.~\ref{Calc_Co_zigzag}). On the other hand, this zigzag structure appears around $V_{\rm zigzag} = 0.85 \
{\rm eV}$ in the experimental STM images (see Section~\ref{subsect:Co_adsorption}). The determination of the angle $\alpha$ in
Fig.~\ref{Calc_Co_zigzag} allows for an easy comparison to Fig.~\ref{Ge2x1+Co_atoms}~(d). It can be seen that, apart from a small energy
mismatch, the Co induced perturbation of the calculated zigzag structure again perfectly matches the experimental observation.

The energy mismatch between the calculated and experimental STM results can be related to the doping of the investigated Ge crystals (\emph{p}-type Ge
crystals with a low dopant concentration are used in this work), which is not included in the DFT modeling. Since the surface and bulk bands shift in
energy depending on the type of doping and on the doping level, it can be expected that the Co induced perturbations shift in energy as well. On the other
hand, whereas deposition of $0.032$ ML of Co did not lead to changes of the Ge(111)2$\times$1 electronic structures in the experiments, the incorporation
of the Co atom in the DFT model may induce a ``doping" effect of the Co/Ge(111)2$\times$1 system due to the finite size of the SC. For the 23$\times$5 SC,
the Co/Ge ratio is 1/2071, which corresponds to a heavily doped Ge crystal. As already mentioned in Section~\ref{subsect:Co_adsorption}, the
precise value of the tunneling voltage $V_{\rm zigzag}$ is found to depend on the semiconductor type and on the doping concentration. E.g., heavily doped
\emph{n}-Ge(111)2$\times$1 surfaces (phosphor doping level $n_{\rm P} = 1 \times 10^{19} \, {\rm cm}^{-3}$, surface preparation as described in
Section~\ref{sect:Inst}) are found to have $V_{\rm zigzag} = 0.55 \, {\rm V}$, which is in good agreement with the calculated tunneling voltage of $0.57 \,
{\rm V}$.

%========================================================%
%========     Fig 13     " Low_doping_Ge "     ==========%
%========================================================%
%, bb=0 0 946 652
\begin{figure}
\includegraphics[width=80.0mm, scale=1.00]{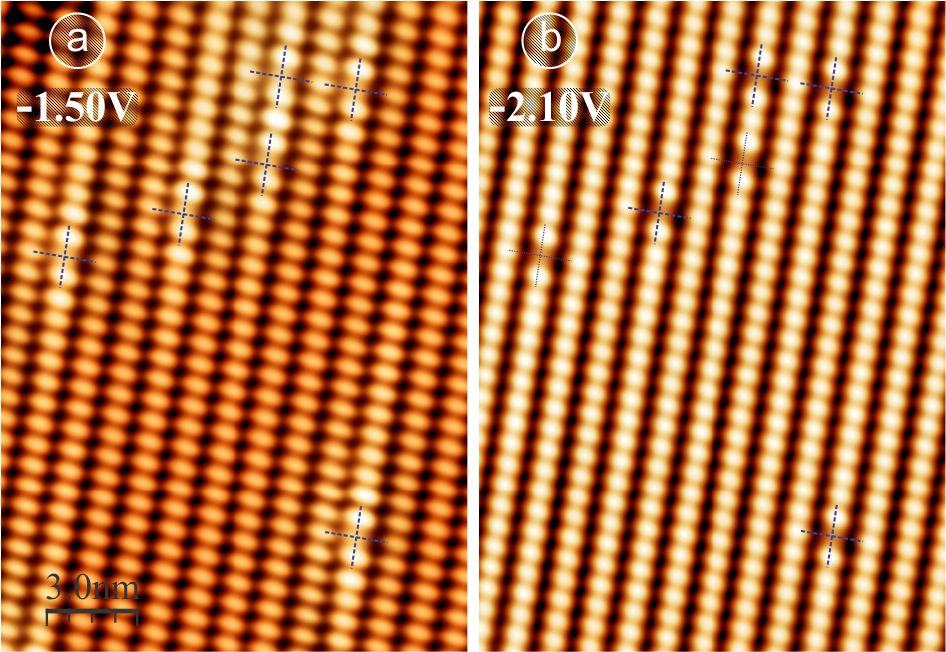}
\caption{(Color online) (a), (b) $14.0 \times 21 \, {\rm nm}^2$ experimental filled states STM images of 6 individual Co atoms for
\emph{n}-type Ge(111)2$\times$1 (resistance $\rho_{\rm bulk} \simeq 11 \, {\rm \Omega~cm}$). The tunneling voltage $V_t$ is indicated for each
image. $I_t$ is fixed at $100 \,{\rm pA}$ for (a) and at $500 \,{\rm pA}$ for (b). Blue cross markers indicate identical locations in (a) and
(b).} \label{Low_doping_Ge}
\end{figure}
%========================================================%
%========================================================%
%========================================================%

Finally, we investigated the influence of the doping level on the appearance of the clean Ge(111)2$\times$1 surface and the Co/Ge(111)2$\times$1 system in
experimental STM topography images. As indicated above, it can be expected that the Co induced perturbations shift in energy, since the surface and bulk
bands shift in energy depending on the type of doping and the doping level. In Fig.~\ref{Low_doping_Ge} we present two experimental filled states
STM topography images of a low doped \emph{n}-type Ge(111)2$\times$1 surface. The Ge crystal was doped by P at a doping level $n_{\rm P} = 0.5$ to $1.0
\times 10^{15} \, {\rm cm}^{-3}$ ($\rho_{\rm bulk} \simeq 11 \, {\rm \Omega~cm}$). Surface preparation and Co deposition are performed as described in
Section~\ref{sect:Inst}. At low temperatures ($T_{\rm sample} \simeq 4.5 \, {\rm K}$), these samples exhibit a detectable tunneling current in the filled
states regime only at tunneling voltages below $-0.8 \ {\rm V}$. Interestingly, the above described Co induced perturbations appear at significantly
different energies for this low doped sample. The features observed in Fig.~\ref{Low_doping_Ge}~(a) for the \emph{n}-type Ge(111)2$\times$1 surface are
observed around $V_{\rm t} = -0.7 \ {\rm V}$ for the \emph{p}-type Ge(111)2$\times$1 surface (see Fig.~\ref{STM_DFT_comparision}), corresponding to an
energy shift of about $0.7 \, {\rm to} \, 0.8 \, {\rm eV}$. A similar energy shift can be inferred by comparison of Fig.~\ref{Low_doping_Ge}~(b) and
Fig.~\ref{STM_DFT_comparision}. The energy shifts between the STM experiments and the DFT calculations in Fig.~\ref{STM_DFT_comparision} may therefore be
attributed to doping effects. Alternatively, the observed energy shifts may also be related to the intrinsic deficiency of LDA with respect to the
quantitative determination of band gap values of semiconductor materials.

%--------------------------------------------------------------------------%
\subsection{Co/Ge(111)2$\times$1 electronic properties and Co--Ge bonding characteristics}
\label{subsect:Co_bonds}
%--------------------------------------------------------------------------%

%========================================================%
%========    TABLE 2   " Co-Ge distances "     ==========%
%========================================================%
\begin{table}[b]
\caption{Calculated distances between the Co atom at site (7) and the neighboring Ge atoms [labeled in Fig.~\ref{Ge_Co_bonds}~(b1)] for the
8$\times$4 SC equilibrium geometry.} \label{Co_Ge_dist}

\begin{ruledtabular}
\begin{tabular}{cccc}
& Co neighboring atoms & Co--Ge distance & \\
\colrule
  & Co--Ge(1) & 2.687 $\rm \AA$ &\\ % top red Co atom
  & Co--Ge(2) & 2.404 $\rm \AA$ &\\ % green Co atom
  & Co--Ge(3) & 2.404 $\rm \AA$ &\\ % green Co atom
  & Co--Ge(4) & 2.415 $\rm \AA$ &\\ % purple  Co atom
  & Co--Ge(5) & 2.243 $\rm \AA$ &\\ % bulk light-blue/green Co atom
\end{tabular}
\end{ruledtabular}
\end{table}
%========================================================%
%========================================================%
%========================================================%

%========================================================%
%========       TABLE 3     " Mulliken "       ==========%
%========================================================%
\begin{table}[b]
\caption{(A) Mulliken overlap population and electron population of the perturbed Ge atoms neighboring the embedded Co atom at site (7) and (B)
of the unperturbed Ge atoms in the absence of the Co atom (ideal Ge(111)2$\times$1 surface). Ge atoms are numbered according to
Fig.~\ref{Ge_Co_bonds}~(b1) and Table~\ref{Co_Ge_dist}.} \label{Ge_Co_Mulliken}

\begin{ruledtabular}
\begin{tabular}{crrrrr}
\textbf{A.} Co/Ge(111)2$\times$1 &  &  &  &  & \\
Bond (atom--atom)& Ge(1) & Ge(2) & Ge(3) & Ge(4) & Ge(5) \\
\colrule
   Ge--Ge   &  0.368 &  0.444 &  0.396 &  0.368 &  0.228 \\
   Ge--Ge   &  0.394 &  0.394 &  0.445 &  0.401 &  0.356 \\
   Ge--Ge   &  0.396 &  0.409 &  0.408 &  0.228 &  0.401 \\
   Ge--Ge   &        &  0.067 &  0.065 &  0.401 &  0.353 \\
   Co--Ge   &  0.120 &  0.241 &  0.239 &  0.224 &  0.231 \\
  Electron pop. &  4.205 &  3.891 &  3.889 &  3.943 &  3.884 \\ %\hline
\colrule
\textbf{B.} Ideal (2$\times$1)~~~~~~ &  &  &  &  & \\
Bond (atom--atom)& Ge(1$'$) & Ge(2$'$) & Ge(3$'$) & Ge(4$'$) & Ge(5$'$) \\   %\hline
\colrule
   Ge--Ge   &  0.412 &  0.475 &  0.475 &  0.412 &  0.395 \\
   Ge--Ge   &  0.476 &  0.476 &  0.456 &  0.402 &  0.406 \\
   Ge--Ge   &  0.475 &  0.435 &  0.435 &  0.395 &  0.405 \\
   Ge--Ge   &    -   &  0.089 &  0.089 &  0.402 &  0.406 \\
  Electron pop. &  4.172 &  3.882 &  3.859 &  3.982 &  3.961 \\
\end{tabular}
\end{ruledtabular}
\end{table}
%========================================================%
%========================================================%
%========================================================%

\begin{figure*}
\includegraphics[width=177mm, scale=1.00]{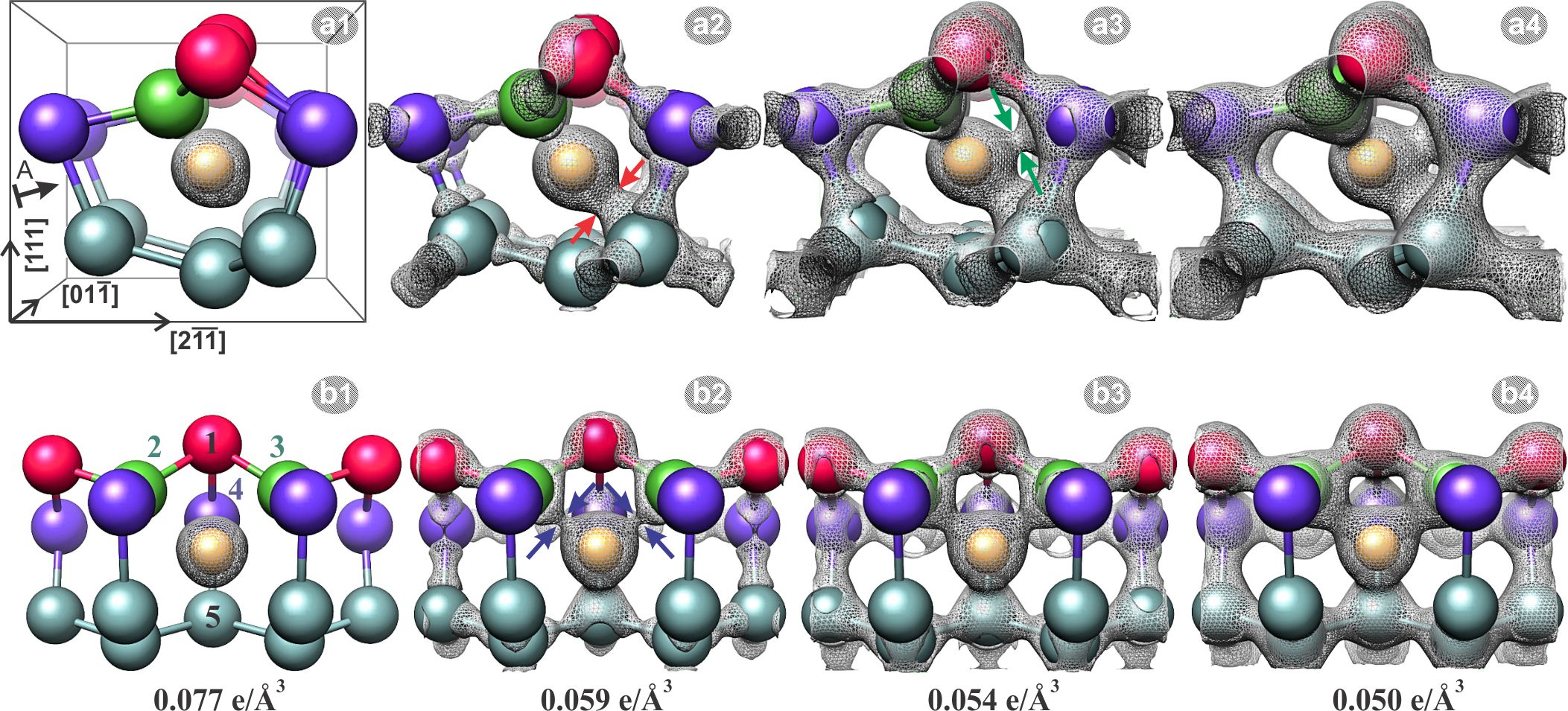}
\caption{(Color online) 3D isosurface charge density $\rho(x,y,z)$ plots of the Co/Ge(111)2$\times$1 system  (chain-left isomer) with the Co
atom located inside the big 7-member Ge ring below the surface. Viewpoint is along the [01$\overline{1}$] direction for (a1) to (a4) and along
the direction indicated by the arrow A in (a1) for (b1) to (b4). The charge density $\rho$ (${\rm e/}$\AA$^3$) isovalues for (a1), (a2), (a3)
and (a4) are the same as the ones for (b1), (b2), (b3) and (b4), respectively, and are indicated at the bottom of the latter parts of the
figure.} \label{Ge_Co_bonds}
\end{figure*}

In this section we discuss the chemical bonds that are formed between the Co atom at site (7) and its surrounding Ge atoms [labeled in
Fig.~\ref{Ge_Co_bonds}~(b1)] underneath the Ge(111)2$\times$1 surface. The amount of possible chemical bonds can be roughly estimated by relying
on a simple analysis of the Co-Ge bond lengths. For a single bond between a Co and a Ge atom the bond length can be estimated as the sum of the
Co and Ge covalent radii, which is $r_{\rm c} = 2.38${\AA}. The calculated distances between the Co atom and the neighboring Ge atoms for the
8$\times$4 SC equilibrium geometry (see subsection~\ref{subsect:Co_sites}) are listed in Table~\ref{Co_Ge_dist}. Among the listed Ge atoms, Ge
atoms (2) to (5) are most likely to form a covalent bond with the Co atom. The distances between the Co atom and these Ge atoms are, however,
slightly larger (1-2\%) than $r_{\rm c}$, which can be attributed to the employed LDA (an overestimation of the bond lengths by a few percent is
typical for LDA). On the other hand, the calculated Co--Ge(1) distance is significantly larger (around 13\%) than $r_{\rm c}$, implying that the
formation of a bond between the Co atom and the Ge(1) \emph{up}-atom can be excluded.

%========================================================%
%===========      Fig 14      " E_cov "       ===========%
%========================================================%
%, bb=0 0 1023 579
\begin{figure}
\includegraphics[width=85.3mm, scale=1.00]{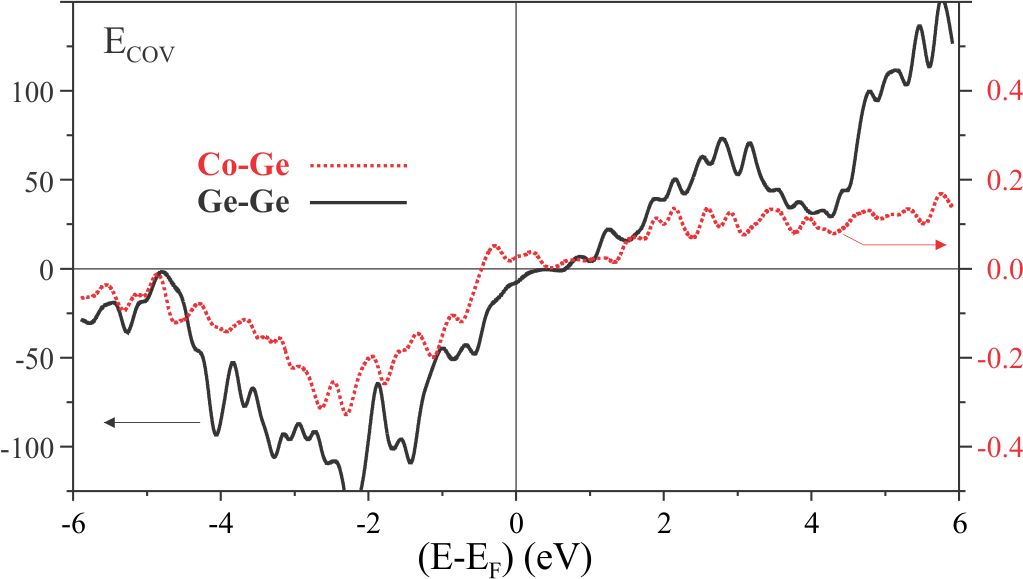}
\caption{(Color online) Chemical bonding in terms of the covalent bond energy ($E_{\rm COV}$) for Ge--Ge bulk and Co--Ge interactions.}
\label{E_cov}
\end{figure}
%========================================================%
%========================================================%
%========================================================%

%========================================================%
%========     Fig 15     " Ge_Co_bonds "       ==========%
%========================================================%
%, bb=0 0 2105 955
%========================================================%
%========================================================%
%========================================================%

An estimate of the atomic charges, electron transfer and covalent interactions between the Co atom at site (7) and the neighboring Ge atoms
[labeled in Fig.~\ref{Ge_Co_bonds}~(b1), also see Table~\ref{Co_Ge_dist}] can be obtained more qualitatively from the calculated number of
electrons inside the atomic spheres and from the Mulliken populations.~\cite{Mulliken_JCP_95} The Mulliken electron orbital overlap populations
(calculated  for the 23$\times$5 SC) of Ge atoms (1) to (5) both with (perturbed $\pi$-bonded chain row of the Co/Ge(111)2$\times$1 system) and
without (unperturbed $\pi$-bonded chain row of the Ge(111)2$\times$1 surface) the Co atom are given in Table~\ref{Ge_Co_Mulliken}~(A) and in
Table~\ref{Ge_Co_Mulliken}(B), respectively. The (unperturbed) bulk atoms Ge(4$'$) and Ge(5$'$) have four Ge--Ge covalent bonds, with an
electron orbital overlap population around $0.40 \pm 0.01 \ {\rm e/bond}$ ($sp^3$ hybridization). On the other hand, \emph{up}-atom Ge(1$'$) and
\emph{down}-atoms Ge(2$'$,3$'$) have only three covalent bonds ($sp^2$ hybridization): A fourth bond is absent or at least strongly reduced
(Mulliken overlap populations are less than 22\% of the Ge--Ge bulk covalent bond). Co-induced changes in the Mulliken overlap population of the
neighboring Ge atoms and the atom population can be traced by comparison of Table~\ref{Ge_Co_Mulliken}~(A) to Table~\ref{Ge_Co_Mulliken}~(B).
The calculations reveal that no significant charge redistribution between the Co atom and the surface and bulk Ge atoms occurs, i.e. the
electron population of the Co atom remains close to that of a neutral Co atom ($9.043 \ {\rm e}$). The Co atom exhibits four covalent bonds with
Mulliken overlap populations that are 56\% to 60\% of the Ge--Ge bulk covalent bond. There exists a weak interaction with the Ge(1)
\emph{up}-atom as well (see Table~\ref{Ge_Co_Mulliken}~(A)). Because of the very low overlap population (30\% of Ge--Ge bulk bond) and the very
weak charge transfer between Co and Ge(1), however, the interaction between the Co and Ge(1) \emph{up}-atom should not be considered as a fifth
covalent bond.

The energy of the covalent Co-Ge bonds can be estimated as the total energy difference between a Co atom that is ``bonded" [Co atom located at
site (7)] or ``not bonded" to the Ge lattice. Calculations for the ``not bonded'' case were performed for a 4$\times$2 SC with the Co atom
placed 4\AA~ above the Ge surface. The total energy difference is found to be $8.4 \pm 0.3 \ {\rm eV}$. Considering four Co--Ge bonds as
discussed above, the (average) energy of a single Co--Ge bond is hence $E_{\rm Co-Ge} = 2.1 \ {\rm eV}$, which is significantly lower than the
energy of the covalent Ge--Ge bond ($E_{\rm Ge-Ge} = 3.71 \ {\rm eV}$).

The analysis of the bond lengths and the Mulliken overlap populations are indicative of rather weak Co--Ge bonds, which can also be concluded from
our experimental observation that Co atoms diffuse even at lower temperatures.~\cite{Muzychenko_Submitted_11} More precisely, in spite of the low sample
temperature ($T_{\rm sample} \, \leq 80 \, {\rm K}$) during Co deposition, the majority (around 87\%) of the Co atoms diffuses along $\pi$-bonded chain
rows to surface/subsurface defects (including DBs, MASs, and subsurface Ga impurities), which can be related to the weak Co--Ge bonds. The remaining 13\%
can be retrieved as individual (subsurface) Co atoms at the cold Ge surface and they diffuse as well to surface/subsurface defects after warming the
substrate up to room temperature. For a more detailed discussion we refer the reader to Ref.~[\cite{Muzychenko_Submitted_11}].

The chemical bonding mechanism between the Co and Ge atoms involved in the DFT electronic structure calculations can be investigated in more
detail by evaluating the crystal orbital overlap population/Hamiltonian population (COOP/COHP).~\cite{Hoffmann_AChem_93, Dronskowski_JPC_93} We
have used an alternative COOP/COHP based approach that allows to calculate the relevant physical quantities independent of the choice of zero of
the potential by relying on the so-called ``covalent bond energy'' ($E_{\rm COV}$).~\cite{Bester_JP_01} COOP and $E_{\rm COV}$ calculations are
known to yield similar results, while the COOP method generally overestimates the magnitude of the anti-bonding states when defined within a
plane-wave basis set.~\cite{Matar_PRB_04} Figure~\ref{E_cov} presents our $E_{\rm COV}$ calculation for the Co--Ge and Ge--Ge bulk interactions
in an energy range of $12 \ {\rm eV}$ around the Fermi level. Note that the $E_{\rm COV}$ values (y-ordinate) are plotted without any units and
can only be interpreted qualitatively. Negative, positive, and zero values of $E_{\rm COV}$ correspond to bonding, anti-bonding, and non-bonding
interactions, respectively. The $E_{\rm COV}$ spectra confirm the stability of the Co/Ge system. Strong bonding interactions exist for Ge--Ge
from the bottom of the VB up to the Fermi level, while for the Co--Ge a strong bonding interaction is found only up to around $-0.3 \ {\rm eV}$.
Between $-0.3 \ {\rm eV}$ and the Fermi level, Co--Ge anti-bonding interactions occur, which may explain the experimentally observed thermal
instability of the Co/Ge system (the covalent Co--Ge bonds are weaker than the Ge--Ge bulk bonds).

Finally, we calculated charge electron density maps to visualize the Co--Ge bonds. In Fig.~\ref{Ge_Co_bonds} we present isosurface maps of the
spatial electron charge density $\rho(x,y,z)$ for isosurface values ranging from $0.077 \, {\rm e/}$\AA$^3$ down to $0.050 \, {\rm e/}$\AA$^3$.
High electron densities related to the Ge--Ge covalent bonds gradually appear above $0.059 \ {\rm e/}$\AA$^3$ in Figs.~\ref{Ge_Co_bonds}~(a2,b2)
to (a4,b4). Moreover, zones of high electron localization between the Co atom and the surrounding Ge atoms can be observed. A zone of high
electron localization exists between the Co atom and the Ge(5) atom in Fig.~\ref{Ge_Co_bonds}~(a2) (indicated by red arrows). Two additional
symmetrical zones of high electron localization can be observed between the Co atom and the Ge(2) and Ge(3) \emph{down}-atoms in
Fig.~\ref{Ge_Co_bonds}~(b2) (indicated by the blue arrows). A fourth Co--Ge bond can be related to the high electron localization zone between
the Co atom and the Ge(4) atom in Fig.~\ref{Ge_Co_bonds}~(a3) (indicated by the green arrows). As could be expected from our above analysis,
there occurs no high electron localization zone between the Co atom and the \emph{up}-atom Ge(1), which additionally confirms that both atoms
are not bonded. The findings related to the bond lengths, the Mulliken overlap populations and the $E_{\rm COV}$ calculations are hence in
agreement with the isosurface charge density maps. We therefore conclude that the Co atom forms a bond with four neighboring Ge atoms,
corresponding to a Co$^{\rm 4+}$ valence state and a 3$d^{\rm 5}$ electron configuration. It is known that Co has a wide range of valence states
due to its various spin configurations, implying Co$^{\rm 4+}, i.e. $ 3$d^{\rm 5}$ ions can exist in several spin
configurations.~\cite{Sen_PRA_92, Ravindran_PRB_99} An extension of our DFT model for the embedding of a Co atom in the Ge lattice by including
spin-dependent calculations will be a topic of future research.

%%%%%%%%%%%%%%%%%%%%%%%%%%%%%%%%%%%%%%%%%%%%%%%%%%%%%%%%%%%%%%%%%%%%%%%%%%%%
%%%%%%%%%%%%%%%%%%%%%%%%%%%%%%%%%%%%%%%%%%%%%%%%%%%%%%%%%%%%%%%%%%%%%%%%%%%%
%%%%%%%%%%%%%%%%%%                Conclusion                %%%%%%%%%%%%%%%%
%%%%%%%%%%%%%%%%%%%%%%%%%%%%%%%%%%%%%%%%%%%%%%%%%%%%%%%%%%%%%%%%%%%%%%%%%%%%
%%%%%%%%%%%%%%%%%%%%%%%%%%%%%%%%%%%%%%%%%%%%%%%%%%%%%%%%%%%%%%%%%%%%%%%%%%%%
\section{Conclusion}
Noninvasive embedding of individual Co atoms into clean Ge(111)2$\times$1 surfaces was systematically investigated by means of STM experiments
and DFT calculations. STM experiments indicate that these Co atoms appear exclusively at upper $\pi$-bonded chain rows after deposition on cold
Ge(111)2$\times$1 surfaces ($T_{\rm sample} \leq 80 \, {\rm K}$). Analysis of the voltage dependent STM images reveals that all adsorbed Co
atoms induce an identical anisotropic electronic perturbation of the surrounding Ge surface and clear 1D confinement along $\pi$-bonded chains.
Relying on DFT based calculations we demonstrated that the energetically most favorable position of a Co atom is attained by penetration into
the Ge(111)2$\times$1 surface. The Co atom occupies a quasi-stationary position within the big 7-member Ge ring of the Ge(111)2$\times$1
reconstruction in between the $3^{\rm rd}$ and $4^{\rm th}$ atomic layer beneath the surface. The embedded Co atom induces an electronic
asymmetry of the $\pi$-bonded chain with respect to the $[01\overline{1}]$ direction, which allows us to determine that the Ge(111)2$\times$1
surface reconstruction of all investigated Ge samples consists of domains with chain-left geometry exclusively. Calculated STM images based on
our DFT approach match very well the experimental STM images of the Co/Ge(111)2$\times$1 system within the investigated range of tunneling
voltages. Finally, DFT based calculations of the Co--Ge bond strength reveal the formation of four covalent bonds, corresponding to a Co$^{\rm
4+}$ valence state and a 3$d^{\rm 5}$ electron configuration. Our findings open interesting perspectives for investigations of subsurface 1D
(nanowires) and 2D (islands) nanostructures that are expected to form at a higher coverage of Co on Ge(111).

\begin{acknowledgments}
The research in Moscow has been supported by the Russian Foundation for Basic Research (RFBR) grants and by the computing facilities of the
M.V.~Lomonosov Moscow State University (MSU) Research Computing Center. The research in Leuven has been supported by the Research Foundation -
Flanders (FWO, Belgium) as well as by the Belgian Interuniversity Attraction Poles (IAP) and the Flemish Concerted Action (GOA) research
programs. K. S. is a postdoctoral researcher of the FWO.
\end{acknowledgments}

%\bibliography{Muzychenko_v09}

\end{document}